\documentclass[aps,prd,superscriptaddress,showpacs,preprint]{revtex4}
\usepackage{graphicx, bm}
\usepackage[usenames]{color}
\usepackage{float}
\usepackage{subcaption}

\begin{document}

\draft
\title{Research on the electromagnetic and weak dipole moments of the tau-lepton at the Bestest Little Higgs Model}

\author{E. Cruz-Albaro\footnote{elicruzalbaro88@gmail.com}}
\affiliation{\small Facultad de F\'{\i}sica, Universidad Aut\'onoma de Zacatecas\\
            Apartado Postal C-580, 98060 Zacatecas, M\'exico.\\}

\author{ A. Guti\'errez-Rodr\'{\i}guez\footnote{alexgu@fisica.uaz.edu.mx}}
\affiliation{\small Facultad de F\'{\i}sica, Universidad Aut\'onoma de Zacatecas\\
         Apartado Postal C-580, 98060 Zacatecas, M\'exico.\\}

\author{J. I. Aranda 
}
\affiliation{\small Facultad de Ciencias F\'{\i}sico Matem\'aticas, Universidad Michoacana de San Nicol\'as de Hidalgo\\
            Avenida Francisco, J. M\'ujica S/N, 58060, Morelia, Michoac\'an, M\'exico.\\}

\author{F. Ram\'irez-Zavaleta 
}
\affiliation{\small Facultad de Ciencias F\'{\i}sico Matem\'aticas, Universidad Michoacana de San Nicol\'as de Hidalgo\\
            Avenida Francisco, J. M\'ujica S/N, 58060, Morelia, Michoac\'an, M\'exico.\\}



\date{\today}

\begin{abstract}

In this paper, using the Bestest Little Higgs Model (BLHM) we calculate at the one-loop level the contributions to the Anomalous Magnetic Dipole Moment (AMDM) and Anomalous Weak Magnetic Dipole Moment (AWMDM) of the tau-lepton. The implications from this model are study, emphasizing the contributions of the new physics induced  by the new scalar and vector bosons of the BLHM: $S_i=H_0, A_0, \phi^{0}, \eta^{0}, \sigma, H^{\pm}, \phi^{\pm}, \eta^{\pm}$, and  $V_i=Z', W'$, because these quantify the new physics. With these new contributions we estimated bounds on both the real and imaginary parts of the AMDM and AWMDM of the tau-lepton.
Our study complements other one-loop level research performed on models beyond the Standard Model.

\end{abstract}


\vspace{5mm}

\maketitle

\section{Introduction}

The study of the physics of the tau-lepton by the ATLAS and CMS experiments~\cite{Neutelings:2021bsl,ATLAS:2017mpa,CMS:2016gvn,ATLAS:2015xbi,ATLAS:2014rzk} at the Large Hadron Collider (LHC) has developed significantly and now represents a very active physics program. In addition, the following present and future colliders: hadron-hadron ($pp$), lepton-hadron ($e^{-}p$) and lepton-lepton ($ e^{+}e^{-},\, \mu^{+} \mu^{-} $) for the post LHC era will open up new horizons in the field of fundamental physics. All of these colliders contemplate in their physics programs the study of the physics of the tau-lepton.

In the Standard Model (SM) of elementary particle physics, as well as in many of its extensions, the search for Anomalous Magnetic Dipole Moments (AMDM), Electric Dipole Moments (EDM) and Anomalous Weak Magnetic Dipole Moments (AWMDM) of fundamental fermions, and in particular from the tau-lepton is an important aspect of theoretical, phenomenological and experimental investigations hunting for physics beyond the Standard Model (BSM) of particle physics. For a review on the bounds on the electromagnetic and weak dipole moments see Refs.~\cite{Gutierrez-Rodriguez:2022eyn,Gutierrez-Rodriguez:2022mtt,Dyndal:2020yen,Koksal:2018vtt,Koksal:2018env,Eidelman:2016aih,Atag:2015xjs,Hayreter:2013vna,Gutierrez-Rodriguez:2013eaa,Gutierrez-Rodriguez:2009weo,Passera:2007fk,Gutierrez-Rodriguez:2006abh,GutierrezRodriguez:2004ch,Pich:2013lsa,ATLAS:2012qtn}.

In the lepton sector, the tau-lepton is  a key particle in the SM and in several extensions of the SM as it is considered as a laboratory for many experimental or simulation aspects of the search for new physics. 
This particle is characterized by its high mass~\cite{Data2020} compared to the mass of the electron or muon, so one would expect its electromagnetic and weak dipole moments to be much more sensitive to the effects of new physics than the electron or muon itself~\cite{Pich:2013lsa}. 
Unfortunately, the very short $\tau $ lifetime~\cite{Data2020} makes very difficult to measure its dipole moments (AMDM, EDM, AWMDM) with a precision good enough to perform a significative test. The spin-precession technique adopted in the electron and muon $g-2$ is no-longer feasible~\cite{Pich:2013lsa}. Instead, one measures the production of tau pairs at different high-energy processes. 
For instance, the most stringent current bound on the $\tau$ AMDM (see Table~\ref{valores-atau}) was derived using the data collected by the DELPHI Collaboration from measurements in the cross-section of the process $e^{+} e^{-} \to e^{+} e^{-} \tau^{+} \tau^{-}$ at $\sqrt{s}$ between 183 and 208 GeV at LEP2~\cite{DELPHI:2003nah}. 
As for the $ \tau $ EDM, $ d_{\tau} $, the BELLE Collaboration searched for CP-violation effects in the $e^{+} e^{-} \to \gamma^{*} \to  \tau^{+} \tau^{-} $ process using triple momentum and spin correlations~\cite{Belle:2002nla}. Through this reaction they obtained the limits shown in Table~\ref{valores-atau} for the real and imaginary parts of the $ \tau $ EDM.
In the SM scenario, the theoretical predictions on the $\tau$ AMDM and EDM are: $  a^{SM}_{\tau}=117721 ( 5) \times 10^{-8}$~\cite{Samuel:1990su,Hamzeh:1996np,Eidelman:2007sb} and $  d^{SM}_{\tau}< 10^{-34}\, e$cm~\cite{Hoogeveen:1990cb,Pospelov:1991zt,Barr-Marciano}, respectively. These results are well below current experimental limits.

\begin{table}[H]
\caption{The best current  experimental results for the electromagnetic dipole moments of the $ \tau $-lepton.
\label{valores-atau}}
\center
\begin{tabular}{c c c c }
\hline
$ \text{Collaboration} $  &  \hspace{0.8cm}   $ \text{Best present experimental bounds on } a_\tau \ \text{and}\ d_\tau$   & \hspace{0.8cm} $\text{C.L.}$  & \hspace{0.8cm}  $\text{Reference}$  \\
\hline
\hline
 DELPHI    & \hspace{0.8cm} $ -0.052 < a_\tau <  0.013$ &  \hspace{0.8cm} $ 95\, \% $  &  \hspace{0.8cm}  \cite{DELPHI:2003nah}  \\
 BELLE    & \hspace{0.8cm} $ -2.2 < \text{Re}(d_\tau (10^{-17}\, e\,cm)) < 4.5 $ &  \hspace{0.8cm} $ 95\, \%   $  &  \hspace{0.8cm} \cite{Belle:2002nla}\\
     & \hspace{0.8cm} $ -2.5 < \text{Im}(d_\tau (10^{-17}\, e\,cm)) < 0.08 $ &  \hspace{0.8cm} $ 95\, \%   $  &  \hspace{0.8cm} \cite{Belle:2002nla}\\
\hline
\end{tabular}
\end{table}

Another intrinsic property of the $ \tau $-lepton that has received  attention in recent years due to important advances in the experimental domain consists of the weak dipole moments of the tau, which are associated with its interaction with the  $Z$ gauge boson. Both the AWMDM and the Weak Electric Dipole Moment (WEDM) of the $ \tau $-lepton, $ a^{W}_{\tau}$ and $d^{W}_{\tau} $, have been investigated with LEP data~\cite{ALEPH:2002kbp,L3:1998lhr,OPAL:1996dwj}. In Table~\ref{valores-awtau} we show the current best experimental bounds on $a^{W}_{\tau}$ and $d^{W}_{\tau} $.   
These limits were obtained through $ \tau^{+} \tau^{-} $ production at LEP by the ALEPH Collaboration, corresponding to an integrated luminosity of 155 $ \text{pb}^{-1} $~\cite{ALEPH:2002kbp}.
On the theoretical side, the reached precisions in the AWMDM and WEDM of the tau-lepton are: $ a^{W-SM}_{\tau}=-(2.10+ 0.61\, \text{i})\times 10^{-6}$~\cite{Bernabeu:1994wh} and  $d^{W-SM}_{\tau}< 8\times 10^{-34}\ e$cm~\cite{Bernreuther:1988jr}. These values are well below the current experimental sensitivity.
 This opens the possibility to look for deviations from the SM and therefore, it is worthwhile to study extensions of the SM as they could generate large contributions of new physics that are closer to the experimental bounds. 
With these motivations, we research on the electromagnetic and weak dipole moments of the tau-lepton in the context of the BLHM.

Based on everything already mentioned above, in this paper, we estimate the sensitivity bounds on the AMDM and AWMDM  of the $ \tau $-lepton in the SM and BLHM scenario, and emphasize will be placed on the contributions generated by the particles predicted by the BLHM, as these quantify the new physics.

\begin{table}[H]
\caption{The best current  experimental results for the weak dipole moments of the $ \tau $-lepton.
\label{valores-awtau}}
\center
\begin{tabular}{c c c c }
\hline
$ \text{Collaboration} $  &  \hspace{0.8cm}   $ \text{Best present experimental bounds on } a^{W}_\tau \ \text{and}\ d^{W}_\tau$   & \hspace{0.8cm} $\text{C.L.}$  & \hspace{0.8cm}  $\text{Reference}$  \\
\hline
\hline
 ALEPH    & \hspace{0.8cm} $ |\text{Re}( a^{W}_\tau)| < 1.14 \times 10^{-3} $ &  \hspace{0.8cm} $ 95\, \% $ &  \hspace{0.8cm}  \cite{ALEPH:2002kbp}  \\
    & \hspace{0.8cm} $ |\text{Im}(a^{W}_{\tau})|< 2.65 \times 10^{-3} $ &  \hspace{0.8cm} $ 95\, \%   $  &  \hspace{0.8cm} \cite{ALEPH:2002kbp}\\
  ALEPH   & \hspace{0.8cm} $ |\text{Re}(d^{W}_\tau) | < 0.50 \times 10^{-17}\, e\,cm$ &  \hspace{0.8cm} $ 95\, \%   $  &  \hspace{0.8cm} \cite{ALEPH:2002kbp}\\
    & \hspace{0.8cm} $  |\text{Im}(d^{W}_{\tau} )|< 1.1 \times 10^{-17}\ e\,cm$ &  \hspace{0.8cm} $ 95\, \%   $  &  \hspace{0.8cm} \cite{ALEPH:2002kbp}\\
\hline
\end{tabular}
\end{table}

The purpose of the BLHM is to solve the hierarchy problem without fine-tuning. This is achieved through the incorporation of one-loop corrections to the Higgs boson mass through heavy top-quarks partners and heavy gauge bosons.
This extension of the SM predicts the existence of new physical scalar bosons neutral and charged $h_0, H_0, A_0, \phi^{0},\eta^{0},
\sigma, H^{\pm}, \phi^{\pm}, \eta^{\pm}$, new heavy gauge bosons $Z', W'$ and new heavy quarks $B, T,T_5, T_6, T^{2/3},T^{5/3}$.
At the one-loop level, the  AMDM $  a_{\tau}$ and  AWMDM $  a^{W}_{\tau}$  of the $ \tau $-lepton are induced via the Feynman diagrams represented in Figs.~\ref{dipolo} and~\ref{dipoloweak}, where $S_i$ represent scalar bosons, $V_i$ neutral and charged gauge bosons and $l_i$ leptons. 
In the framework of the BLHM, new model contributions are those arising from the vertices of scalars bosons and vector bosons, that is to say, vertices of the form  (see Figs.~\ref{dipolo} and~\ref{dipoloweak}):  $\tau l_iS_i$, $S_i= H_0, A_0, \phi^{0}, \eta^{0}, \sigma, H^{\pm}, \phi^{\pm},
\eta^{\pm}$,  $\tau l_iV_i$, $V_i= Z', W'$, and $Zl_i\bar l_i$, $l_i=\tau, \nu_{\tau}$,
respectively. With these vertices we calculate the one-loop contributions to the 
 AMDM and AWMDM of the $ \tau $-lepton and in several scenarios with $m_{A_0}= 1000$ GeV, $m_{\eta^0}= 100$ GeV, $\tan\beta=3$, $f=[1000, 3000]$ GeV and $F=[3000,6000]$ GeV.

The paper is structured as follows. In Section II, we give a brief review of the BLHM. In Section III, we present the predictions of the BLHM on the electromagnetic and weak dipole moments of the tau-lepton. 
In Section IV, we discuss the sensitivity bounds obtained on the AMDM and  AWMDM of the $ \tau $-lepton.
 Finally, we present our conclusions in Section V.
In Appendix A, we present the set of Feynman rules employed in the study of electromagnetic and weak dipole moments of the $ \tau $-lepton in the context of the BLHM. 
In Appendix B, we provide the one-loop level SM predictions on the AMDM and AWMDM of the tau-lepton.

\section{The Bestest Little Higgs Model}

Various extensions of the SM, such as Little Higgs Models (LHM)~\cite{Arkani1,Arkani2}, have been proposed in order to solve the problem of the mass hierarchy.  This class of models employs a complex mechanism named collective symmetry breaking. Its main
idea is to represent the SM Higgs boson as a pseudo-Nambu-Goldstone boson of an approximate global symmetry which is spontaneously broken at a scale in the TeV range.
In these models, the collective symmetry breaking mechanisms is implemented in the norm sector, fermion sector and the Higgs
sector, which predict new particles within the mass range of a few TeV. These new particles play the role of partners of the
top-quark, of the gauge bosons and the Higgs boson, the effect of which is to generate radiative corrections for the mass
of the Higgs boson, and thus cancel the divergent corrections induced by SM particles.
However, LHM~\cite{Arkani1,Arkani2,Arkani3} are already strongly constrained by electroweak precision data. 
These constraints typically require the new gauge bosons of LHM to be quite heavy \cite{PRD67-2003,PRD68-2003}. In most LHM, the top partners are heavier than the new gauge bosons, and this can lead to significant fine-tuning
in the Higgs potential~\cite{JHEP03-2005}.

An interesting and relatively recent model is the BLHM \cite{JHEP09-2010} overcomes these difficulties by including separate symmetry breaking scales at which the heavy gauge boson and top partners obtain their masses. This model  generates heavy gauge boson partner masses above the already excluded mass range, and has relatively light top partners below the upper bound from fine-tuning.
The BLHM is based on two independent non-linear sigma models. With the first field $\Sigma$, the global symmetry
$SO(6)_A\times SO(6)_B$ is broken to the diagonal group $SO(6)_V$ at the energy scale $f$, while with the second field
$\Delta$, the global symmetry $SU(2)_C \times SU(2)_D$ to the diagonal subgroup $SU(2)$ to the scale $F> f$. In the first
stage are generated 15 pseudo-Nambu-Goldstone bosons that are parameterized as

\begin{equation}\label{Sigma}
\Sigma=e^{i\Pi/f}  e^{2i\Pi_{h}/f}e^{i\Pi/f},
\end{equation}

\noindent 
where $\Pi$ and $\Pi_h$ are complex and antisymmetric matrices given in Ref.~\cite{JHEP09-2010}. Regarding the second stage of spontaneous symmetry-breaking, the pseudo-Nambu-Goldstone bosons of the field $\Delta$ are parameterized as follows

\begin{equation}\label{Delta}
\Delta=F e^{2i \Pi_d/F},\, \,\, \, \, \Pi_d=\chi_a \frac{\tau^{a}}{2} \ \ (a=1,2,3),
\end{equation}

\noindent 
$\chi_a$ represents the Nambu-Goldstone fields and the $\tau_a$ correspond to the Pauli matrices~\cite{JHEP09-2010}, which are the
generators of the SU(2) group.

\subsection{The scalar sector}

The BLHM Higgs fields, $h_1$ and $h_2$, form the Higgs potential that undergoes spontaneous symmetry breaking~\cite{JHEP09-2010,Kalyniak,Erikson}:

\begin{equation}\label{Vhiggs}
V_{Higgs}=\frac{1}{2}m_{1}^{2}h^{T}_{1}h_1 + \frac{1}{2}m_{2}^{2}h^{T}_{2}h_2 -B_\mu h^{T}_{1} h_2 + \frac{\lambda_{0}}{2} (h^{T}_{1}h_2)^{2}.
\end{equation}

\noindent The potential reaches a minimum when $m_1, m_2 >0$, while to break the electroweak symmetry requires $B_\mu > m_1 m_2$. The symmetry-breaking mechanism is implemented in the BLHM when the Higgs doublets acquire
their vacuum expectation values (VEVs), $\langle h_1\rangle ^{T}=(v_1,0,0,0)$ and $ \langle h_2 \rangle ^{T}=(v_2,0,0,0)$. By demanding that
these VEVs minimize the Higgs potential of Eq.~(\ref{Vhiggs}), the following relations are obtained

\begin{eqnarray}\label{v12}
&&v^{2}_1=\frac{1}{\lambda_0}\frac{m_2}{m_1}(B_\mu-m_1 m_2),\\
&&v^{2}_2=\frac{1}{\lambda_0}\frac{m_1}{m_2}(B_\mu-m_1 m_2).
\end{eqnarray}

\noindent These parameters can be expressed as follows

\begin{equation}\label{vvacio}
v^{2}\equiv v^{2}_1 +v^{2}_2= \frac{1}{\lambda_0}\left( \frac{m^{2}_1 + m^{2}_2}{m_1 m_2} \right) \left(B_\mu - m_1 m_2\right)\simeq \left(246\ \ \text{GeV}\right)^{2},
\end{equation}

\begin{equation}\label{beta}
\text{tan}\, \beta=\frac{v_1}{v_2}=\frac{m_2}{m_1}.
\end{equation}

\noindent   
From the diagonalization of the mass matrix for the scalar sector,
three non-physical fields $G_0$ and $G^{\pm}$, two physical scalar fields $H^{\pm}$
and three neutral physical scalar fields $h_0$, $H_0$ and $A_0$ are generated~\cite{Kalyniak,PhenomenologyBLH}. The lightest state, $h_0$, is identified as the
scalar boson of the SM. The masses of these fields are given as

\begin{eqnarray}\label{masaAGH}
m_{G_0}&=&m_{G^{\pm}}=0,\\
m^{2}_{A_{0}}&=&m^{2}_{H^{\pm}} =m^{2}_1+m^{2}_2,\label{mHmas} \\
m^{2}_{h_0,H_{0}} &=& \frac{B_\mu}{\text{sin}\, 2\beta}\mp \sqrt{\frac{B^{2}_{\mu}}{\text{sin}^{2}\, 2\beta} -2\lambda_0 B_\mu v^{2} \text{sin}\, 2\beta +\lambda^{2}_{0} v^{4} \text{sin}^{2}\, 2\beta  } \label{mh0H0}.
\end{eqnarray}

\noindent
  The four parameters present in the Higgs potential $ m_1,  m_2, B_\mu$ and $\lambda_0 $, can be replaced by another more phenomenologically accessible set. That is, the masses of the states $h_0$ and $A_0$, the angle $\beta$ and the VEV $v$~\cite{Kalyniak}:

\begin{eqnarray}\label{parametros}
B_\mu &=&\frac{1}{2}(\lambda_0  v^{2} + m^{2}_{A_{0}}  )\, \text{sin}\, 2\beta,\\
\lambda_0 &=& \frac{m^{2}_{h_{0}}}{v^{2}}\Big(\frac{  m^{2}_{h_{0}}- m^{2}_{A_{0}} }{m^{2}_{h_{0}}-m^{2}_{A_{0}} \text{sin}^{2}\, 2\beta }\Big),\\
\text{tan}\, \alpha &=& \frac{ B_\mu \text{cot}\, 2\beta+ \sqrt{(B^{2}_\mu/\text{sin}^{2}\, 2\beta)-2\lambda_0 B_\mu v^{2} \text{sin}\, 2\beta+ \lambda^{2}_{0} v^{4}\text{sin}^{2}\, 2\beta  }  }{B_\mu -\lambda_0 v^{2} \text{sin}\, 2\beta},\label{alpha}   \\
m^{2}_{H_{0}} &=& \frac{B_\mu}{\text{sin}\, 2\beta}+ \sqrt{\frac{B^{2}_{\mu}}{\text{sin}^{2}\, 2\beta} -2\lambda_0 B_\mu v^{2} \text{sin}\, 2\beta +\lambda^{2}_{0} v^{4} \text{sin}^{2}\, 2\beta  }, \label{mH0}\\
m^{2}_{\sigma}&=&(\lambda_{56} + \lambda_{65})f^{2}=2\lambda_0 f^{2} \text{K}_\sigma. \label{masaescalar}
\end{eqnarray}

\noindent 
 The variables $\lambda_{56}$ and $\lambda_{65}$ in Eq.~(\ref{masaescalar}) represent the coefficients of the quartic potential defined
in~\cite{JHEP09-2010}, both variables take values different from zero to achieve the collective breaking of the symmetry
and generate a quartic coupling of the Higgs boson~\cite{JHEP09-2010,Kalyniak}.
The BLHM also contains scalar triplet fields that get a contribution to their mass from the explicit symmetry breaking terms in model, as define in Ref.~\cite{JHEP09-2010}, that depends on the parameter $m_4$.

\begin{eqnarray}
m^{2}_{\phi^{0}}&=& \frac{16}{3}F^{2} \frac{3 g^{2}_{A} g^{2}_{B}}{32 \pi^{2}} \log \left( \frac{\Lambda^{2}}{m^{2}_{W'}}\right) + m^{2}_{4} \frac{f^{4}+ F^{4}}{F^{2}(f^{2}+F^{2})},\\
m^{2}_{\phi^{\pm}}&=& \frac{16}{3}F^{2} \frac{3 g^{2}_{A} g^{2}_{B}}{32 \pi^{2}} \log \left( \frac{\Lambda^{2}}{m^{2}_{W'}}\right) + m^{2}_{4} \frac{f^{4}+f^{2}F^{2}+F^{4}}{F^{2}(f^{2}+F^{2})},\\
m^{2}_{\eta^{\pm}}&=&  m^{2}_{4}+ \frac{3 f^{2} g^{2}_{Y}}{64 \pi^{2}}\frac{\Lambda^{2}}{F^{2}},\\
m^{2}_{\eta^{0}}&=&m^{2}_{4}. 
\end{eqnarray}

\subsection{The gauge sector}

In the BLHM the new gauge bosons develop masses proportional to $\sqrt{f^2+F^2}\sim F$. This makes the masses of the gauge bosons large relative to other particles that have masses proportional to $f$. The kinetic terms of the gauge fields in the BLHM
are given as follows:

\begin{equation}\label{Lcinetico}
\mathcal{L}=\frac{f^{2}}{8} \text{Tr}(D_{\mu} \Sigma^{\dagger} D^{\mu} \Sigma) + \frac{F^{2}}{4} \text{Tr}(D_\mu \Delta^{\dagger} D^{\mu} \Delta),
\end{equation}

\noindent where

\begin{eqnarray}\label{derivadasC}
D_{\mu}\Sigma&=&\partial_{\mu} \Sigma +i g_A A^{a}_{1\mu} T^{a}_L \Sigma- i g_B \Sigma A^{a}_{2\mu} T^{a}_L+ i g_{Y} B^{3}_{\mu}(T^{3}_{R}\Sigma-\Sigma T^{3}_{R}),\\
D_{\mu}\Delta&=&\partial_{\mu} \Delta +i g_A A^{a}_{1\mu} \frac{\tau^{a}}{2}  \Delta- i g_B \Delta A^{a}_{2\mu} \frac{\tau^{a}}{2}.
\end{eqnarray}

\noindent 
 $T^{a}_{L}$ are the generators of the group $SO(6)_A$ corresponding to the subgroup $SU(2)_{LA}$, while $T^3_R$ represents
the third component of the $SO(6)_B$ generators corresponding to the $SU(2)_{LB} $ subgroup, these matrices are provided in~\cite{JHEP09-2010}.
$g_A$ and $A^{a}_{1\mu}$ denote the gauge coupling and field associated with the gauge bosons of $SU(2)_{LA}$. $g_B$ and $A^{a}_{2\mu}$
represent the gauge coupling and the field associated with $SU(2)_{LB}$, while $g_Y$ and $B^{3}_{\mu}$ denote the hypercharge and the field.
When $\Sigma$ and $\Delta$ get their VEVs, the gauge fields $A^{a}_{1\mu}$ and $A^{a}_{2\mu}$ are mixed to form a massless triplet
$A^{a}_{0\mu}$ and a massive triplet $A^{a}_{H\mu}$,

\begin{equation}\label{AA}
A^{a}_{0\mu}=\text{cos}\, \theta_g A^{a}_{1\mu} + \text{sin}\, \theta_g A^{a}_{2\mu}, \hspace{5mm} A^{a}_{H\mu}= \text{sin}\, \theta_g A^{a}_{1\mu}- \text{cos}\, \theta_g A^{a}_{2\mu},
\end{equation}

\noindent with the mixing angles

\begin{equation}\label{gagb}
s_g\equiv \sin \theta_g=\frac{g_A}{\sqrt{g_{A}^{2}+g_{B}^{2}} },\ \ c_g \equiv \cos \theta_g=\frac{g_B}{\sqrt{g_{A}^{2}+g_{B}^{2}} },
\end{equation}

\noindent 
which are related to the electroweak gauge coupling $g$ through

\begin{equation}\label{g}
\frac{1}{g^{2}}=\frac{1}{g^{2}_A}+\frac{1}{g^{2}_B}.
\end{equation}

After the breaking of the electroweak symmetry, when the Higgs doublets, $h_1$ and $h_2$ acquire their VEVs, the masses
of the gauge bosons of the BLHM are generated. In terms of the model parameters, the masses are given by

\begin{eqnarray}\label{masaBoson}
m^{2}_{\gamma} &=0&, \\
m^{2}_{Z}&=&\frac{1}{4}\left(g^{2}+g^{2}_Y \right)v^{2} \left(1-\frac{v^{2}}{12 f^2} \left(2+\frac{3f^2}{f^2+F^2} \left( s^{2}_g -c^{2}_g \right)^{2} \right)  \right), \\
m^{2}_{W}&=& \frac{1}{4} g^{2} v^{2} \left(1-  \frac{v^{2}}{12 f^2} \left(2+  \frac{3f^2}{f^2+F^2} \left(s^{2}_g -c^{2}_g  \right)^{2}\right)  \right),\\
m^{2}_{Z'}&=&m^{2}_{W'} +  \frac{g^2 s^{2}_W v^4}{16 c^{2}_W (f^2+F^2)} \left(s^{2}_g -c^{2}_g \right)^{2}, \label{mzprima} \\
m^{2}_{W'}&=& \frac{g^2}{4 c^{2}_{g} s^{2}_{g}} \left(f^2+F^2 \right)  - m^{2}_{W}. \label{mwprima}
\end{eqnarray}

The weak mixing angle is defined as

\begin{eqnarray}\label{angulodebil}
s_W&&\equiv\sin \theta_W = \frac{g_Y}{\sqrt{g^2+ g^{2}_Y }}, \\
c_W&&\equiv\cos \theta_W= \frac{g}{\sqrt{g^2+ g^{2}_Y }},\\
x_W&=&\frac{1}{2 c_W} s_g c_g (s^{2}_g -c^{2}_g).
\end{eqnarray}

\subsection{The fermion sector} \label{subsecfermion}

To construct the Yukawa interactions in the BLHM, the fermions must be transformed under the group $SO(6)_A$ or $SO(6)_B$.
In this model, the fermion sector is divided into two parts. First, the sector of massive fermions is represented by Eq.~(\ref{Ltop}).
This sector includes the top and bottom quarks of the SM and a series of new heavy quarks arranged in four multiplets, $Q$, and $Q'$
which transform under $SO(6)_A$, while $U^c$ and $U^{'c}_5$  are transformed under the group $SO(6)_B$. Second, the sector of
light fermions contained in  Eq.~(\ref{Lligeros}), in this expression all the interactions of the remaining fermions of the SM with the
exotic particles of the BLHM are generated.

For massive fermions, the Lagrangian that describes them is given by~\cite{JHEP09-2010}
\begin{equation}\label{Ltop}
\mathcal{L}_t=y_1 f Q^{T} S \Sigma S U^{c} + y_2 f Q'^{T} \Sigma U^{c} +y_3 f Q^{T} \Sigma U'^{c}_{5} +y_b f q_{3}^{T}(-2 i T^{2}_{R} \Sigma) U^{c}_{b}+ H.c.,
\end{equation}

\noindent  where $ S = \text{diag} (1,1,1,1, -1, -1) $. The explicit representation of the multiplets involved in Eq.~(\ref{Ltop}) are provided in Ref.~\cite{JHEP09-2010,PhenomenologyBLH}. For simplicity, the Yukawa couplings are assumed to be real $y_1, y_2, y_3$ $\in R$.

For light fermions the corresponding Lagrangian is~\cite{JHEP09-2010,PhenomenologyBLH,Martin:2012kqb}
\begin{equation}\label{Lligeros}
\mathcal{L}_{light}= \sum_{i=1,2} y_u f q^{T}_i \Sigma u^{c}_{i} + \sum_{i=1,2} y_{d} f q^{T}_{i}(-2i T^{2}_{R} \Sigma) d^{c}_i
+\sum_{i=1,2,3} y_e f l^{T}_i (-2i T^{2}_{R} \Sigma) e^{c}_i + h.c.
\end{equation}

%
%

\subsection{The currents sector}

The Lagrangian that describes the interactions of fermions with the gauge bosons is~\cite{JHEP09-2010,PhenomenologyBLH}
\begin{eqnarray}\label{LbaseW}
 \mathcal{L} &=& \bar{Q} \bar{\tau}^{\mu} D_{\mu}Q + \bar{Q}' \bar{\tau}^{\mu} D_{\mu}Q'- U^{c\dagger} \tau^{\mu} D_{\mu}U^{c}-  U'^{c\dagger} \tau^{\mu} D_{\mu}U'^{c} -  U_{b}^{c\dagger} \tau^{\mu} D_{\mu}U_{b}^{c} +\sum_{i=1,2}  q^{\dagger}_i \tau^{\mu} D_{\mu} q_i   \nonumber \\
 &+& \sum_{i=1,2,3}  l^{\dagger}_i \tau^{\mu} D_{\mu} l_i
 - \sum_{i=1,2,3}  e_i^{c\dagger} \tau^{\mu} D_{\mu} e^{c}_i - \sum_{i=1,2}  u_{i}^{c\dagger} \tau^{\mu} D_{\mu} u^{c}_{i} - \sum_{i=1,2}  d_{i}^{c\dagger} \tau^{\mu} D_{\mu} d^{c}_i,
\end{eqnarray}

\noindent  where $\tau^{\mu}$ and $\bar{\tau}^{\mu}$ are defined according to~\cite{Spremier}. On the other hand, the respective covariant derivatives are provided in Refs.~\cite{PhenomenologyBLH,Martin:2012kqb}.

\section{Electromagnetic and weak dipole moments of the tau-lepton in the BLHM}


 The electroweak properties of fermions are characterized by physical magnitudes called form factors. These measure properties such as the electric charge, the AMDM, the EDM, the AWMDM, the WEDM and others. Some of these quantities are already present in classical theory, while others arise for the first time as a quantum fluctuation of one-loop or higher orders.
In quantum field theory, the electromagnetic and weak properties of fermions arise through their interaction with the gauge boson $V$, $V= \gamma, Z$.
The most general Lorentz-invariant vertex function describing the interaction of a gauge boson with two fermions can be written in terms of ten form factors \cite{NPB551-1999,NPB812-2009}, which are functions of the kinematic invariants.
In the low energy limit, these correspond to couplings that multiply dimension-four or-five operators in an effective
Lagrangian, and may be complex. If the gauge boson $V$ is on-shell, or if $V$ couples to effectively massless fermions, the number of independent form
factors is reduced to eight. In addition, if the fermions are on-shell, the number is further reduced to four. In this way, the $V \bar{f} f$ vertex function can be written in the form


{\small
\begin{eqnarray}\label{verticeZtt}
ie\bar{u}(p') \Gamma^{\mu}_{ V\bar{f}f} u(p)=ie \bar{u}(p')\big\{ \gamma^{\mu}\left[ F^{V}_{V}(q^{2})-F^{V}_{A}(q^{2})\gamma^{5}\right] 
+ i \sigma^{\mu \nu} q_{\nu} \left[ F^{V}_{M}(q^{2})- iF^{V}_{E}(q^{2})\gamma^{5}\right] \big\}u(p),
\end{eqnarray}
}

\noindent 
where $e$ is the proton charge and $q=p'-p$ the $ V $ gauge boson transferred four-momentum. The terms $F^{V}_{V}(0)$ and $F^{V}_{A}(0)$ in the low energy
limit are the $V\bar{f}f$ vector and axial-vector form factors in the SM, while $F^{V}_M(q^2)$ and $F^{V}_E(q^2)$ are associated with the form factors of the  electromagnetic or weak dipole moments. The latter arise at the loop level and are a valuable tool to study the effects
of new physics indirectly way, through virtual corrections of new particles predicted by extensions of the SM. 
The AWMDM and WEDM are given by  $a^{W}_{f}=-2m_{f}  F^{Z}_M (q^2=m^{2}_{Z})$ and $d^{W}_{f}=- e F^{Z}_E(q^2=m^{2}_{Z})$, whereas the electromagnetic properties, $a_{f}$ and $d_{f}$, are defined by analogue expressions but with the replacement $ q^2=0 $.

\subsection{The AMDM and AWMDM of the tau-lepton at the BLHM}

\begin{figure}[H]
\center
\subfloat[]{\includegraphics[width=4.5cm]{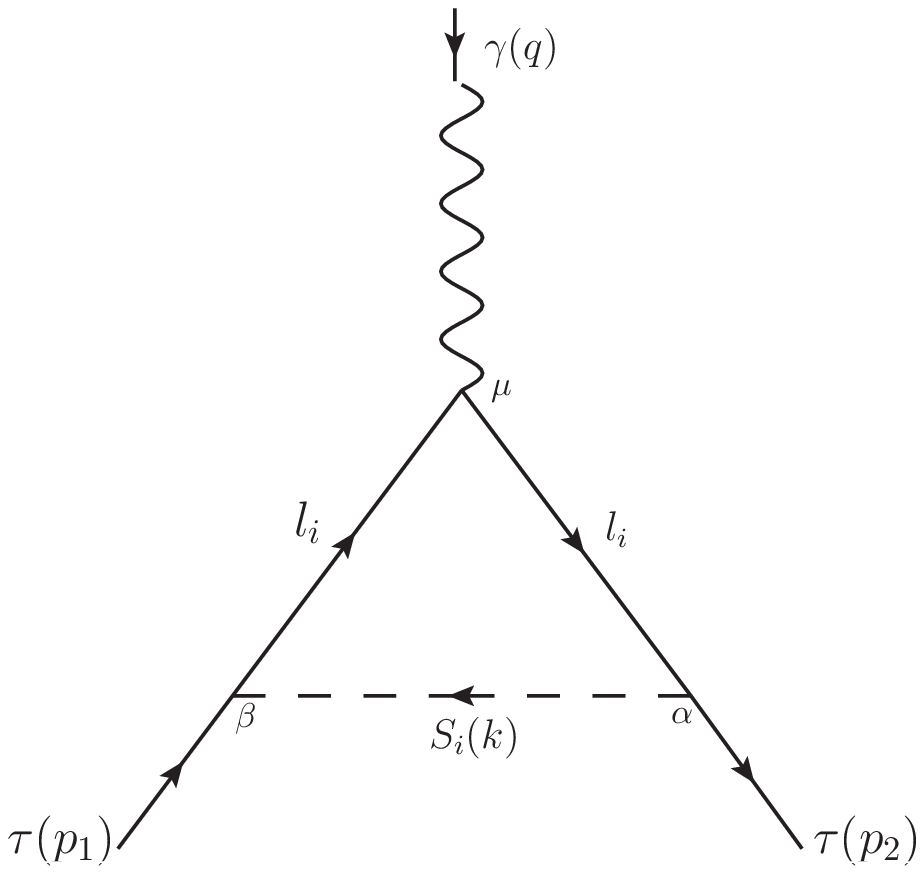}}
\subfloat[]{\includegraphics[width=4.5cm]{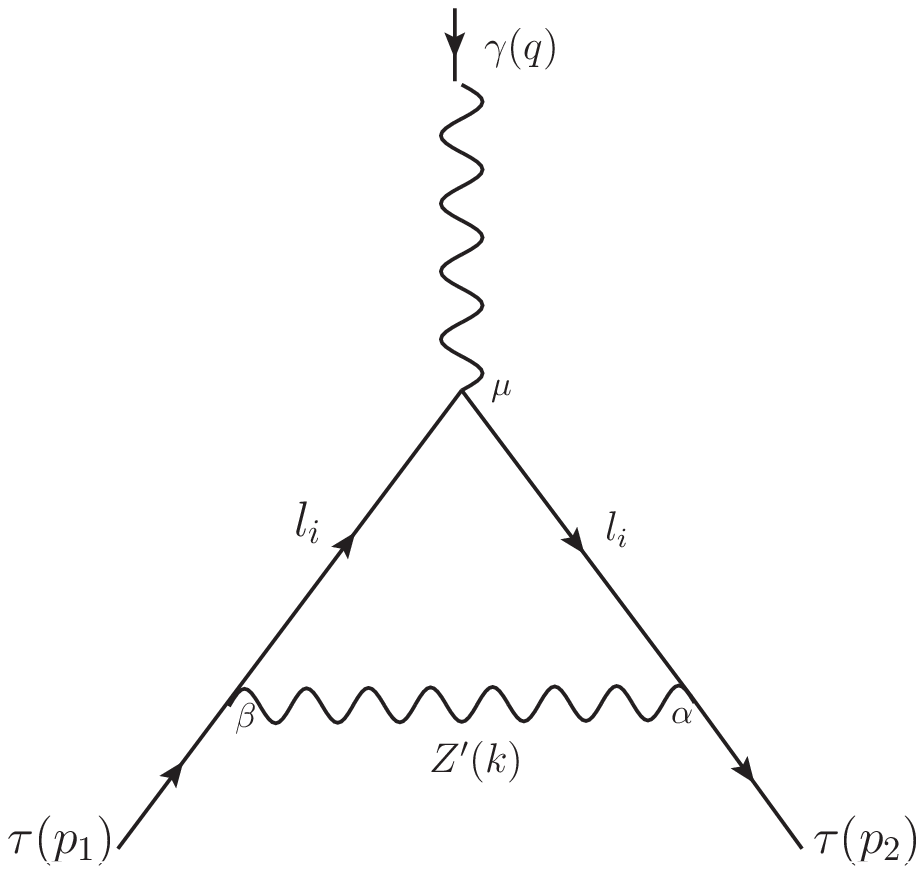}}
\caption{ \label{dipolo} Generic Feynman diagrams that contributes to the AMDM of the tau-lepton, $l_{i}\equiv \tau, \nu_{\tau}$.
 a) Scalar contributions, $S_{i}\equiv \sigma, A_{0}, H_{0}, \eta^{0},\phi^{0}$. b) Vector contribution, $V_i \equiv Z'$.}
\end{figure}

\begin{figure}[H]
\center
\subfloat[]{\includegraphics[width=4.5cm]{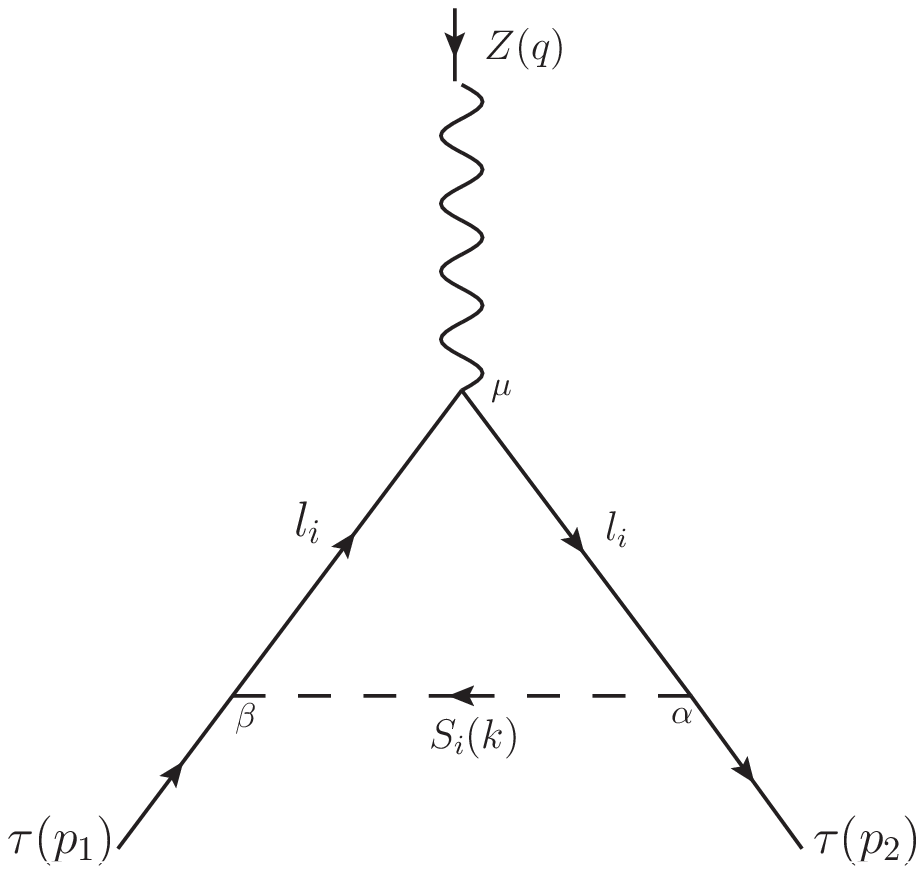}}
\subfloat[]{\includegraphics[width=4.5cm]{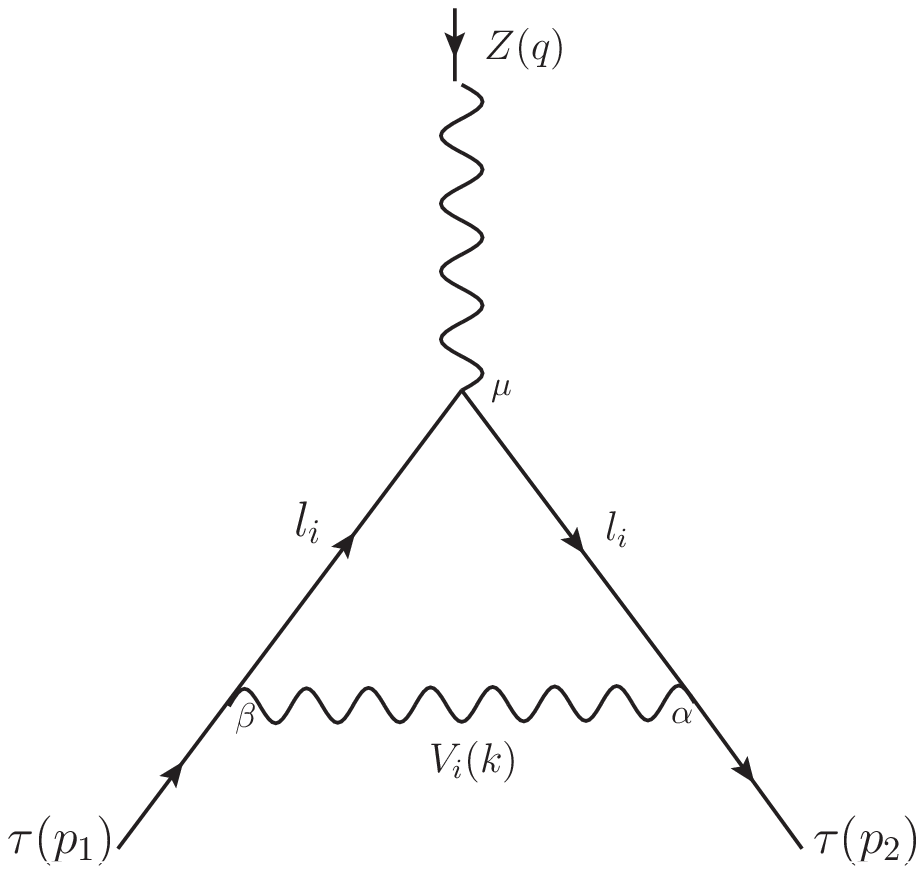}}
\caption{ \label{dipoloweak} Generic Feynman diagrams that contributes to the AWMDM of the tau-lepton, $l_{i}\equiv \tau, \nu_{\tau}$.
 a) Scalar contributions, $S_{i}\equiv A_{0}, H_{0}, H^{\pm},\eta^{0},\phi^{0}, \sigma, \eta^{\pm}, \phi^{\pm}$. b) Vector contributions, $V_{i} \equiv Z', W'$.}
\end{figure}

In this subsection we are interested in the contributions generated by the new BLHM particles to the electromagnetic and weak dipole moments of the tau-lepton. At the one-loop level, the $ \tau $  EDM and WEDM are absent so they do not receive contributions of the radiative corrections. However, the $\tau$  AMDM and AWMDM  are induced by a scalar boson $S_i$ or vector $V_i$, and a pair of $l_{i}$ leptons  via the Feynman diagrams depicted in Figs.~\ref{dipolo} and~\ref{dipoloweak}.
In these figures, $S_{i}$ represents the new scalars $ A_0, H_0, H^{\pm},  \eta^{0},\phi^{0}, \sigma, \eta^{\pm}, \phi^{\pm}$; $V_{i}$ stands for the new gauge
bosons $Z', W'$; and $l_i$ denotes the leptons $\tau, \nu_{\tau}$.
To obtain the amplitude of each contribution we use the Feynman rules provided in Appendix A~\cite{Cruz-Albaro:2022kty}. We used the unitary  gauge for our calculations, and implemented the Passarino-Veltman reduction scheme to solve the loop integrals  involved in the amplitudes.
Such amplitudes are also gauge independent since the $ V $ gauge boson is in on-shell, as well as the tau-lepton pair.

According to Figs.~\ref{dipolo} and~\ref{dipoloweak}, all possible amplitudes contributing to the $F^{\gamma}_M(q^2)$ or  $F^{Z}_M (q^2)$ form factors  can be classified in terms of the two classes of triangle diagrams.
Each category can be written in the following compact notation

\begin{eqnarray}\label{amplitudesSV1}
\mathcal{M}^{\mu}_{\tau}(S_{i})&=& C^{2}_{ \tau l_{i}S_{i}} \int \frac{d^{4}k}{(2\pi)^{4}} \bar{u}(p_{2}) \left[i \frac{\not\! k + \not\!p_{2}+m_{l_i}  }{(k+p_{2})^{2}-m^{2}_{l_{i}}} \right] \left( \gamma^{\mu} (F_{V_i}+F_{A_{i}}\gamma^{5}) \right)  \nonumber \\
&\times & \left[i \frac{ \not\! k + \not\!p_{1}+m_{l_i}  }{(k+p_{1})^{2}-m^{2}_{l_{i}}} \right] u(p_{1}) \left(\frac{i}{k^{2}- m^{2}_{S_{i}}} \right), \\
\mathcal{M}^{\mu}_{\tau}(V_{i}) &=&  \int \frac{d^{4}k}{(2\pi)^{4}} \bar{u}(p_{2}) \left(f^{*}_{V_{i}} \gamma^{\alpha} \right) \left[i \frac{\not\! k + \not\! p_{2}+m_{l_i}  }{(k+p_{2})^{2}-m^{2}_{l_{i}}} \right]  \left( \gamma^{\mu} (F_{V_{i}}+F_{A_{i}}\gamma^{5}) \right) \nonumber \\
&\times & \left[i \frac{ \not\! k + \not\! p_{1}+m_{l_i}  }{(k+p_{1})^{2}-m^{2}_{l_{i}}} \right] \left(f_{V_{i}}\gamma^{\beta} \right) u(p_{1})  \left[\frac{i}{k^{2}- m^{2}_{V_{i}}} \left(-g_{\alpha \beta}+  \frac{k_{\alpha}k_{\beta} }{m^{2}_{V_{i}}}\right) \right],\label{amplitudesSV2} 
\end{eqnarray}

\noindent
 where $F_{V_{i}}$ ($f_{V_{i}}$) and $ F_{A_{i}}$ denote the form factors of the  vector and axial-vector. 
 From these amplitudes we obtain the new physics contributions that are induced by the scalar and vector bosons, particles of the BLHM. The effects of the new physics will be determined in the following way

\begin{eqnarray}\label{awt}
a_{\tau }\equiv a_{\tau}^{BLHM} &=& [a_{\tau}]^{S_i} + [a_{\tau}]^{V_i},\\
a^{W}_{\tau }\equiv a^{W-BLHM}_{\tau} &=& [a^{W}_{\tau}]^{S_i} + [a^{W}_{\tau}]^{V_i}.
\end{eqnarray}

We also consider the total contributions, that is to say, which result from the sum of the contributions of the  SM (see Appendix B) and BLHM.

\begin{eqnarray}
\alpha_{\tau }&=& a_{\tau}^{SM} + a_{\tau}^{BLHM}\label{atau-total},\\
\alpha^{W}_{\tau }&=& a^{W-SM}_{\tau}+ a^{W-BLHM}_{\tau} \label{awtau-total}.
\end{eqnarray}

\section{Numerical results}

For our numerical analysis of the electromagnetic and weak properties of the tau-lepton in the context of the SM and BLHM, we briefly review the free parameters of the BLHM. Subsequently, we discuss the numerical contributions generated for the AMDM and AWMDM of the $\tau$-lepton in each study scenario.

\subsection{Parameters space of the BLHM}

We consider the following BLHM input parameters: $m_{A_{0}}$, $m_{\eta_{0}}$, $\tan \beta$, $\tan \theta_{g}$, $f$ and $F$.

\noindent \textbf{ The pseudoscalar mass $A_{0}$}: This parameter is fixed around 1000 GeV, 
our choice is consistent with the current search results for new scalar bosons~\cite{ATLAS:2020gxx}. Data recorded by the ATLAS experiment at the LHC, corresponding to an integrated luminosity of 139 $\text{fb}^{-1}$ from proton-proton collisions at a centre-of-mass energy 13 TeV, were used to search for a heavy Higgs boson, $A_{0}$, decaying into $ZH$, where $H$ denotes another heavy Higgs boson with mass $m_{H}>125$ GeV.
\vspace{0.1cm}

\noindent   \textbf{ The scalar mass $\eta_{0}$}: In the BLHM scenario, the free parameters $m_{4, 5, 6}$~\cite{JHEP09-2010} are
introduced to break all the axial symmetries in the Higgs potential, giving positive masses to all scalars. Specifically,
the $\eta_{0}$ scalar receives a mass equal to $m_{4}=m_{\eta_{0}}=100$ GeV, according to the BLHM, and the restriction
$m_{4}\gtrsim 10$ GeV must be considered~\cite{JHEP09-2010}. 
\vspace{0.1cm}

\noindent  \textbf{ The ratio of the VEVs of the two Higgs doublets, $\tan \beta$}:
There exists a number of theoretical constraints that can be applied to this parameter, primarily due to perturbativity requirements. The value of $\tan \beta$ is limited by two constraints, the first of which is the requirement that $\lambda_{0}< 4 \pi$, leading to an upper bound according to Eq.~(\ref{cotabeta}). A lower bound also exits, and is set by examinig the radiatively induced contributions to $ m_{1} $ and $ m_{2} $ in the model, which suggest that $\tan \beta>1$~\cite{JHEP09-2010}.

\begin{eqnarray}\label{cotabeta}
 1 < \text{tan}\ \beta < \sqrt{ \frac{2+2 \sqrt{\big(1-\frac{m^{2}_{h_0} }{m^{2}_{A_0}} \big) \big(1-\frac{m^{2}_{h_0} }{4 \pi v^{2}}\big) } }{ \frac{m^{2}_{h_0}}{m^{2}_{A_0}} \big(1+ \frac{m^{2}_{A_0}- m^{2}_{h_0}}{4 \pi v^{2}}  \big) } -1 }.
\end{eqnarray}

\noindent
From this inequality, we can find the range of allowed values for the parameter $\tan \beta$. In particular, for $m_{A_{0}}=1000$ GeV, it is obtained that $1 < \tan \beta < 10.45$. 
\vspace{0.1cm}

\noindent \textbf{ The mixing angle $\theta_{g}$}:
The gauge couplings $g_{A}$ and  $g_{B}$, associated with the $SU(2)_{LA}$ and $SU(2)_{LB}$ gauge bosons,
can be parametrized in a more phenomenological fashion in terms of a mixing angle $\theta_{g}$ and the  $SU(2)_{L}$ gauge coupling:
$\tan \theta_{g}=g_{A}/g_{B}$  and $g=g_{A} g_{B}/\sqrt{g^{2}_{A}+ g^{2}_{B}} $. For simplicity, we can assume that $\tan \theta_{g}=1$,
which implies that the gauge couplings  $g_{A}$ and  $g_{B}$ are equal. The $g_{A,B}$ values are generated using the restriction $g=0.6525$.
\vspace{0.1cm}

\noindent \textbf{ Symmetry breaking scale $f$}:
The BLHM features a global $  SO(6)_{A}\times SO(6)_{B} $ symmetry that is broken to a diagonal $ SO(6)_{V} $ at a scale $f\sim \mathcal{O} $(TeV) when a nonlinear sigma field, $\sum$, develop a VEV. Bounds on the $f$ scale arise when $\tan \beta$ limits, fine-tuning constraints on the heavy quark masses and experimental restrictions  from the production of heavy quarks are taken into account. Refs.~\cite{Kalyniak} and~\cite{Godfrey:2012tf}  establish that $f\in (700,3000)$ GeV.
\vspace{0.1cm}

\noindent \textbf{ Symmetry breaking scale $F$}:
A second global symmetry of the $SU(2)_{C}\times SU(2)_{D}$ form is also present in the BLHM, and is broken to a diagonal $SU(2)$ at a scale $F>f$ when a second nonlinear sigma field, $\Delta$, develops a VEV.
Due to the characteristics of the BLHM, the energy scale $F$ acquires sufficiently large values compared to the $f$ scale. The purpose is to ensure that the new gauge bosons are much heavier than the exotic quarks. In this way, $F\in [3000,6000]$ GeV~\cite{JHEP09-2010,Kalyniak}.

In order to predict the estimates of the AMDM and AWMDM of the tau-lepton, in Table~\ref{parametervalues}, we summarize the values used for the parameters involved in our analysis.

\begin{table}[H]
\caption{Values assigned to the free parameters involved in our numerical analysis in the context of the BLHM.
\label{parametervalues}}
\centering
\begin{tabular}{|c | c|}
\hline
$ \textbf{Parameter} $  & \hspace{1.5cm}    $ \textbf{Value} $  \\
\hline
\hline
$ m_{A_{0}}  $  & \hspace{1.5cm}  $ 1000\ \text{GeV} $  \\
\hline
 $m_{\eta^{0}}  $  & \hspace{1.5cm}   $ 100\ \text{GeV} $  \\
\hline
$ \tan \beta $  & \hspace{1.5cm}   $ 3 $  \\
\hline
$ \tan \theta_{g} $  & \hspace{1.5cm}   $ 1 $  \\
\hline
$ f $  & \hspace{1.5cm}  $ [1000, 3000]\   \text{GeV} $  \\
\hline
$ F $  & \hspace{1.5cm}  $ [3000, 6000] \ \text{GeV} $  \\
\hline
\end{tabular}
\end{table}

\subsection{AMDM of the tau-lepton at the BLHM}

At the one-loop level, the electromagnetic properties of the tau-lepton are induced by the scalar and vector bosons of the BLHM via the Feynman diagrams of Fig.~\ref{dipolo}. 
Below, we focus on the potential effects of the new particles that contribute to the AMDM of tau-lepton, as they could generate a significant enhancement in the value of $ a_{\tau}$ ($a_{\tau }\equiv a_{\tau}^{BLHM}$) compared to the SM prediction $a^{SM}_{\tau}$. As already mentioned, in the BLHM, as well as in the SM, the $\tau$ EDM is multiloop suppressed. Therefore, in this subsection we report only the values of $ a_{\tau}$.

\begin{figure}[H]
\center
\subfloat[]{\includegraphics[width=8.0cm]{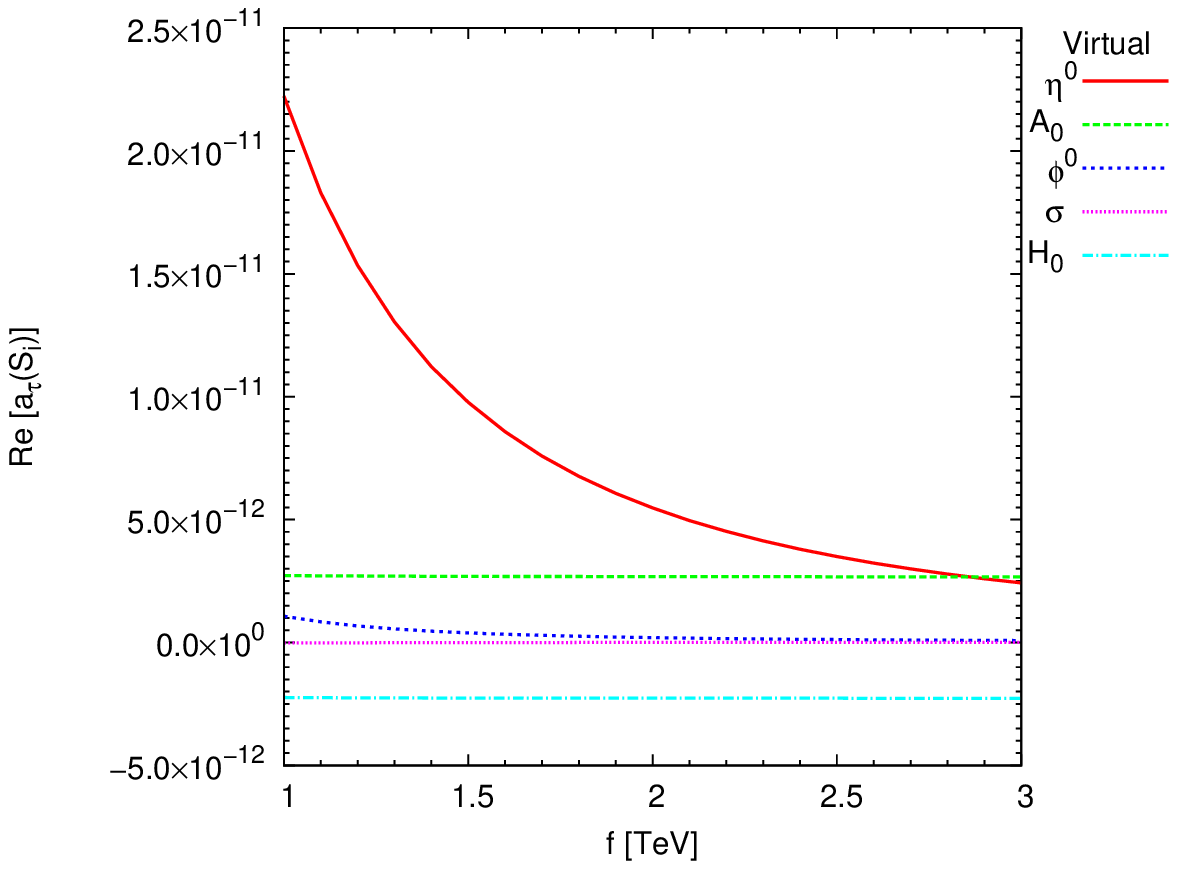}}
\subfloat[]{\includegraphics[width=8.0cm]{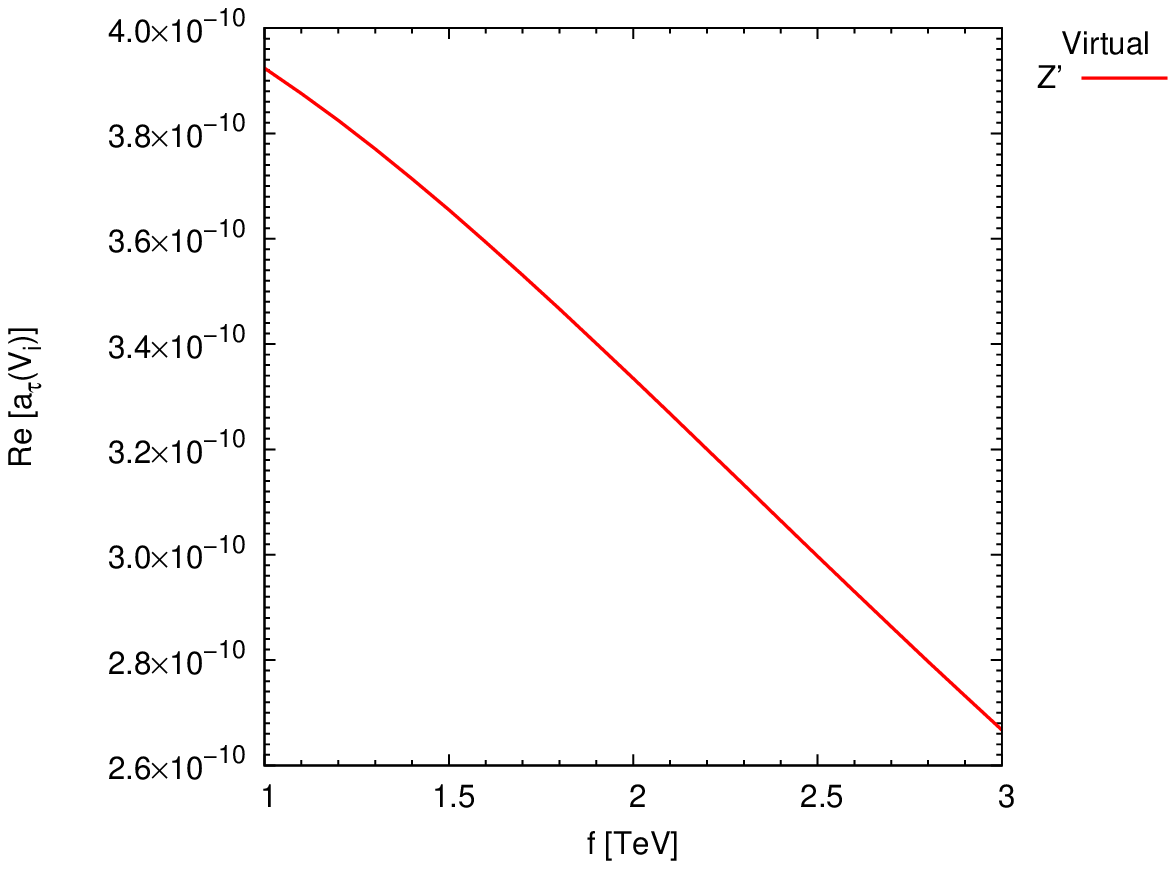}}
\caption{ \label{Si} a) Individual scalar contributions to  Re($ a_{\tau}$). 
b) Individual vector contribution to  Re($ a_{\tau}$). The plots are obtained with the fixed value of $F=4000$ GeV. For the remaining model parameters, the values provided in Table~\ref{parametervalues} are used.}
\end{figure}

For this purpose, we start by solving the amplitudes generated by Eqs.~(\ref{amplitudesSV1}) and~(\ref{amplitudesSV2}), the method we use to solve is the Passarino-Veltman reduction scheme.
From these amplitudes, we extract the form factors proportional to the $\sigma^{\mu \nu}q_{\nu}$ tensor, these form factors contain in coded away the $ [a_{\tau}]^{S_i} $ and  $ [a_{\tau}]^{V_i} $. Thus, we obtain the contributions of each  of the scalar and vector bosons to the AMDM $ a_{\tau}$ of the lepton $ \tau $.
In Fig.~\ref{Si}, we show the partial contributions to $ a_{\tau}$ due to the different particles involved, these individual contributions classified in $ a_{\tau} (S_i)$ and $ a_{\tau} (V_i)$, depend on the energy scale $f$ and generate purely real values.
Specifically, Fig.~\ref{Si}(a) shows the curves of the contributions generated by the scalars $ \eta^{0} $, $ A_{0} $, $ \phi^{0} $, $ \sigma $ and $ H_{0} $.
In this figure, we can appreciate that the scalar $ \eta^{0}$ provides the largest positive contribution with  Re$[a_{\tau}(\eta^0)]$ $=[2.22\times 10^{-11}, 2.43\times 10^{-12}]$ 
while the smallest negative contribution is given by the $H_0$ scalar with Re$[a_{\tau}(H_0)]$ $=-[2.24, 2.27]\times 10^{-12}$.
The remaining scalars generate suppressed contributions, one or more orders of magnitude smaller compared to the main contribution Re$[a_{\tau}(\eta^0)]$: Re$[a_{\tau}(A_0)]$ $ =[2.72,2.67]\times 10^{-12} $, Re$[a_{\tau}(\phi^0)]$ $ =[1.07 \times 10^{-12}, 7.82 \times 10^{-14}]$ and Re$[a_{\tau}(\sigma)]$ $ =[2.09 \times 10^{-14}, 2.96\times 10^{-16} ]$.
On the other hand, the only vector contribution arises from the $Z'$ gauge boson (see Fig.~\ref{Si}(b)). So for the range of analysis set for the  symmetry breaking scale $f$, the contribution of the $Z'$ gauge boson is Re$[a_{\tau}(Z')]$ $ =[3.92, 2.67] \times 10^{-10}$.
According to Figs.~\ref{Si}(a) and~\ref{Si}(b), we observe that the main partial contribution to Re$[ a_{\tau}]$ is generated by the $Z'$ gauge boson. We also notice that as the energy scale $f$ takes values closer to 3000 GeV, the values of Re$[ a_{\tau}]$ become smaller and smaller.
In Table~\ref{parcial} we show the magnitudes of all partial contributions to $  a_{\tau} $ that correspond to the virtual particles circulating in the $\gamma \tau^{+} \tau^{-}$ vertex loop.

\begin{figure}[H]
\center
{\includegraphics[width=8.0cm]{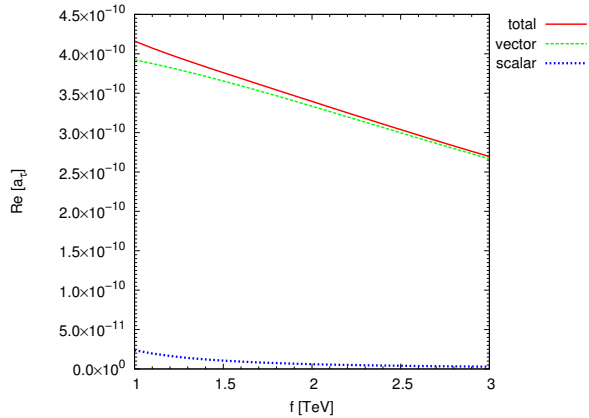}}
\caption{ \label{SVT} Scalar, vector and total contributions to Re$( a_{\tau})$. The plot is obtained with the fixed value of  $F=4000$ GeV. For the remaining model parameters, the values provided in Table~\ref{parametervalues} are used.}
\end{figure}

In Fig.~\ref{SVT} we also describe the behavior of  Re$[ a_{\tau}(S)]$ and  Re$[ a_{\tau}(V)]$, as well as the sum total of these two contributions. Re$[ a_{\tau}(S)]$ and  Re$[ a_{\tau}(V)]$ stand for the sum of all individual contributions to Re$[ a_{\tau}]$ due to the scalar and vector bosons, respectively. 
 Note that the magnitude of the vector contribution dominates with respect to the scalar contribution, so that the total contribution receives significant contributions from the vector sector. The numerical estimates obtained for the three sectors are Re$[ a_{\tau}(V)]$ $=[3.92,2.67]\times 10^{-10}$, Re$[ a_{\tau}(S)]$ $=[2.38\times 10^{-11},2.90\times 10^{-12}]$ and Re$[ a_{\tau}(S+V)]$ $=[4.16,2.70]\times 10^{-10}$ for $f=[1000,3000]$ GeV.
According to these numerical data, we find that effectively the vector and total contribution acquire values of the same order of magnitude, which does not occur with the scalar contribution, which generates slightly small contributions, thus interfering very weakly in the total contribution.

\begin{figure}[H]
\center
\subfloat[]{\includegraphics[width=8.0cm]{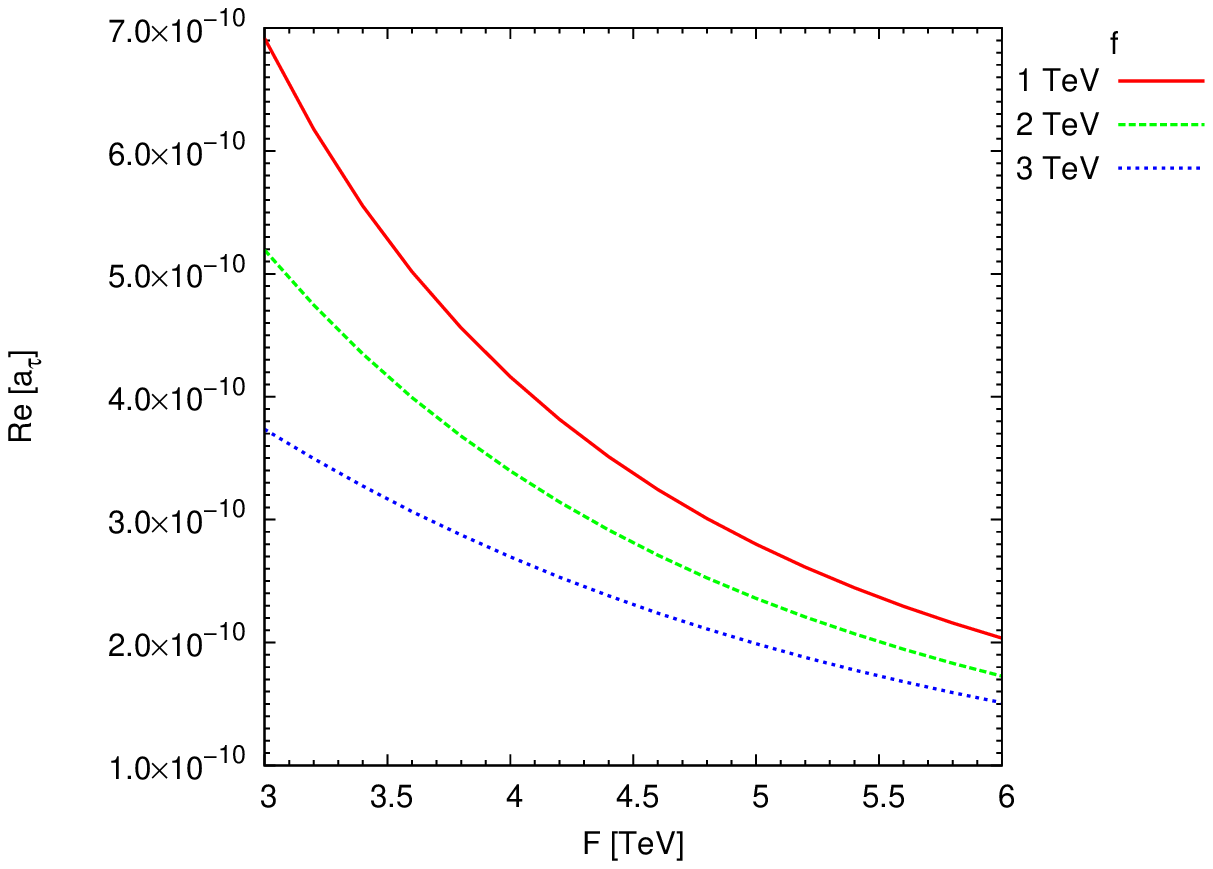}}
\subfloat[]{\includegraphics[width=8.0cm]{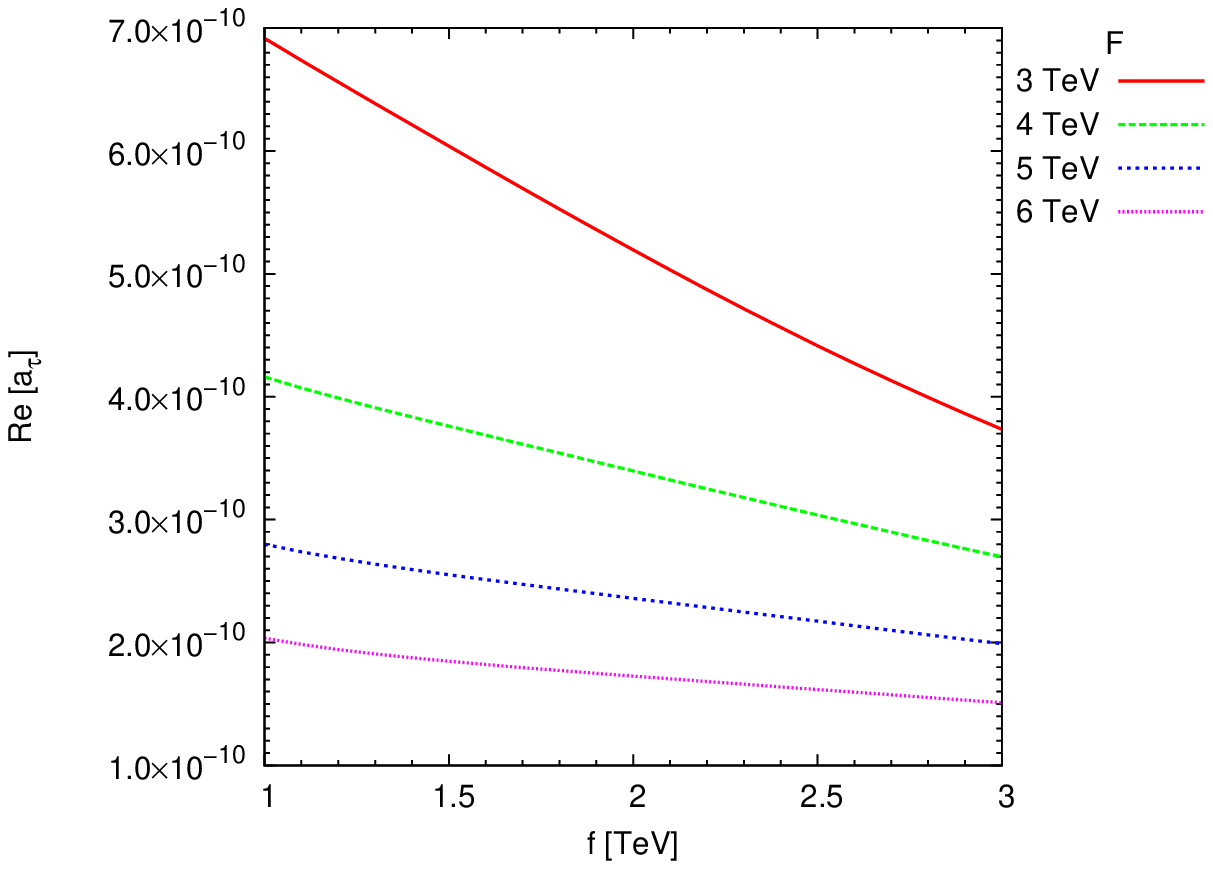}}
\caption{ \label{Fi}  a) Total contribution to Re$( a_{\tau})$ for different values of the energy scale $f$. 
b) Total contribution to Re$( a_{\tau})$ for different values of the energy scale $F$.
The plots are obtained for certain fixed values of the $f$ or $F$ scale. For the remaining parameters of the model, the values provided in Table~\ref{parametervalues} are used.
}
\end{figure}

Until now, the sensitivity of the total AMDM of the tau-lepton has been measured by varying the first symmetry breaking scale $f$. However, it also has dependence on the second symmetry breaking scale $F$. 
Therefore, it is worthwhile to examine the dependence of $ a_{\tau} $ on the scale $F$. Thus, in Fig.~\ref{Fi}(a) we show the level of sensitivity exhibited by the $\tau$ AMDM when varying the $F$ energy scale while keeping the $f$ scale fixed, the fixed values assigned to the $f$ scale are 1000, 2000 and 3000 GeV.
For the three distinct energy scales, we find that the numerical predictions in the $\tau$ AMDM are Re$[ a_{\tau}]$ $=[6.92, 2.04]\times 10^{-10}$, Re$[ a_{\tau}]$ $=[5.19, 1.73]\times 10^{-10}$ and Re$[ a_{\tau}]$ $=[3.74, 1.51]\times 10^{-10}$, respectively.
It is important to note that these contributions to $ a_{\tau} $ acquire only real values and are all of the same order of magnitude, $10^{-10}$, these values do not decrease drastically as the scale of $F$ increases up to 6000 GeV. As we observed in the plot, the dominant contribution arises for small values of the $f$ scale, in particular, when $f=1000$ GeV. 
 With respect to Fig.~\ref{Fi}(b), we plot the curves of the contributions to $ a_{\tau} $ in the analysis range of $f=[1000, 3000]$ GeV, now we fix the scale $F$ and assign values such as 3000, 4000, 5000 and 6000 GeV.
 For these fixed values of $F$, we explore the sensitivity of Re$(a_{\tau})$ and find that the corresponding numerical estimates are Re$[ a_{\tau}]$ $=[6.92, 3.74]\times 10^{-10}$, Re$[ a_{\tau}]$ $=[4.16, 2.70]\times 10^{-10}$, Re$[ a_{\tau}]$ $=[2.80, 1.99]\times 10^{-10}$ and Re$[ a_{\tau}]$ $=[2.04, 1.51]\times 10^{-10}$.
 Again, Re$[ a_{\tau}] \sim 10^{-10}$ and also acquires slightly larger values for small values of the $F$ scale, this being $F=3000$ GeV.
 By way of comparison, we find that Re$[ a_{\tau}]$  takes values of the same order of magnitude if $F$ is varied while $f$ is fixed or the opposite.
Although specifically, Re$[ a_{\tau}]$  obtains slightly larger values when $f=1000$ GeV or $F=3000$ GeV.

\begin{figure}[H]
\center
\subfloat[]{\includegraphics[width=8.0cm]{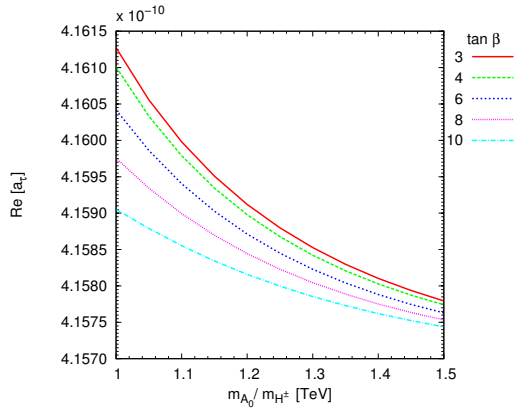}}
\caption{ \label{beta} Total contribution to Re$ (a_{\tau})$ for different values of $\tan \beta$.
The plot is obtained with fixed values of the $f=1000$ GeV and  $F=4000$ GeV. For the remaining parameters of the model, the values provided in Table~\ref{parametervalues} are used.
}
\end{figure}

We now turn to examine the behavior of the real part of $a_{\tau}$ as a function of the mass of the pseudoscalar $A_0$ or the charged scalar $H^{\pm}$, which by the particular characteristics of the BLHM  $m_{A_{0}}=m_{H^{\pm}}$. In this case, we are interested in investigating the phenomenological details associated with the increase of $m_{A_{0}}$ (or $m_{H^{\pm}}$) vs. Re$[a_{\tau}]$.
 According to Eq.~(\ref{cotabeta}), the parameter $\tan \beta$ is directly related to $m_{A_{0}}$, since the range of values that $\tan \beta$ could take is established precisely by the value assigned to $m_{A_{0}}$, i.e., for $m_{A_{0}}=1000$ GeV and $m_{A_{0}}=1500$ GeV the respective ranges of values for the parameter $\tan \beta$, $\tan \beta \in (1,10.45)$  and $\tan \beta \in (1,11.99)$  are generated.
In order to evaluated the numerical contributions of Re$[a_{\tau}]$, we propose to vary the $m_{A_{0}}$ parameter from 1000 GeV to 1500 GeV and also take certain values of $\tan \beta$ in the allowed value space, that is, $\tan \beta =3, 4, 6, 8$ and 10. Fig.~\ref{beta} shows the dependence of Re$[a_{\tau}]$ on $m_{A_{0}}$, we observe that the main signal is reached for  $\tan \beta =3$ while the lowest signal is obtained for  $\tan \beta=10$, 
Re$[ a_{\tau}]=[4.161,4.158]\times 10^{-10}$ and Re$[ a_{\tau}]=[4.159,4.157]\times 10^{-10}$, respectively. For the remaining curves, Re$[ a_{\tau}] \sim 10^{-10}$.
According to our predictions, Re$[ a_{\tau}]$ shows a dependence on the $m_{A_{0}}$ parameter. However, Re$[ a_{\tau}]$ has a small sensitivity to changes in the parameter $ \tan \beta $ since the numerical values obtained by Re$[ a_{\tau}]$ are of the same order of magnitude for different choices in the values of $ \tan \beta $.

\begin{table}[H]
\caption{The magnitude of the partial contributions to $a_{\tau}$ of the BLHM.
The data are obtained by fixing the $f$ and $F$ scales, $f=1000$ GeV and $F=4000$ GeV. For the rest of the model parameters, the values provided in Table~\ref{parametervalues} are used. {\bf abc} denotes the different particles running in the loop of the vertex $\gamma \tau^{+} \tau^{-}$.
\label{parcial}}
\centering
\begin{tabular}{|c|c|}
\hline
\hline
\multicolumn{2}{|c|}{ $f=1 000\ \text{GeV}$, $F=4000\ \text{GeV}$}\\
\hline
 $\text{Couplings}\  \textbf{abc}$  & $\left(a_\tau \right)^{ \textbf{abc} } $  \\
\hline
\hline
$\sigma \tau \tau $  & $- 2_\cdot 09\times 10^{-14} + 0\, i$  \\
\hline
$A_{0} \tau \tau  $  & $ 2_\cdot 72\times 10^{-12} + 0\, i$  \\
\hline
$H_{0} \tau \tau  $  & $ -2_\cdot 24\times 10^{-12}  + 0\ i$  \\
\hline
$\eta^{0} \tau \tau  $  & $ 2_\cdot 22 \times 10^{-11}  + 0\ i$  \\
\hline
$\phi^{0} \tau \tau  $  & $ 1_\cdot 07 \times 10^{-12}  + 0\ i$  \\
\hline
$Z' \tau \tau  $  & $ 3_\cdot 92 \times 10^{-10} + 0\, i$  \\
\hline
\end{tabular}
\end{table}

\subsection{AWMDM of the tau-lepton at the BLHM}

\begin{figure}[H]
\center
\subfloat[]{\includegraphics[width=8.0cm]{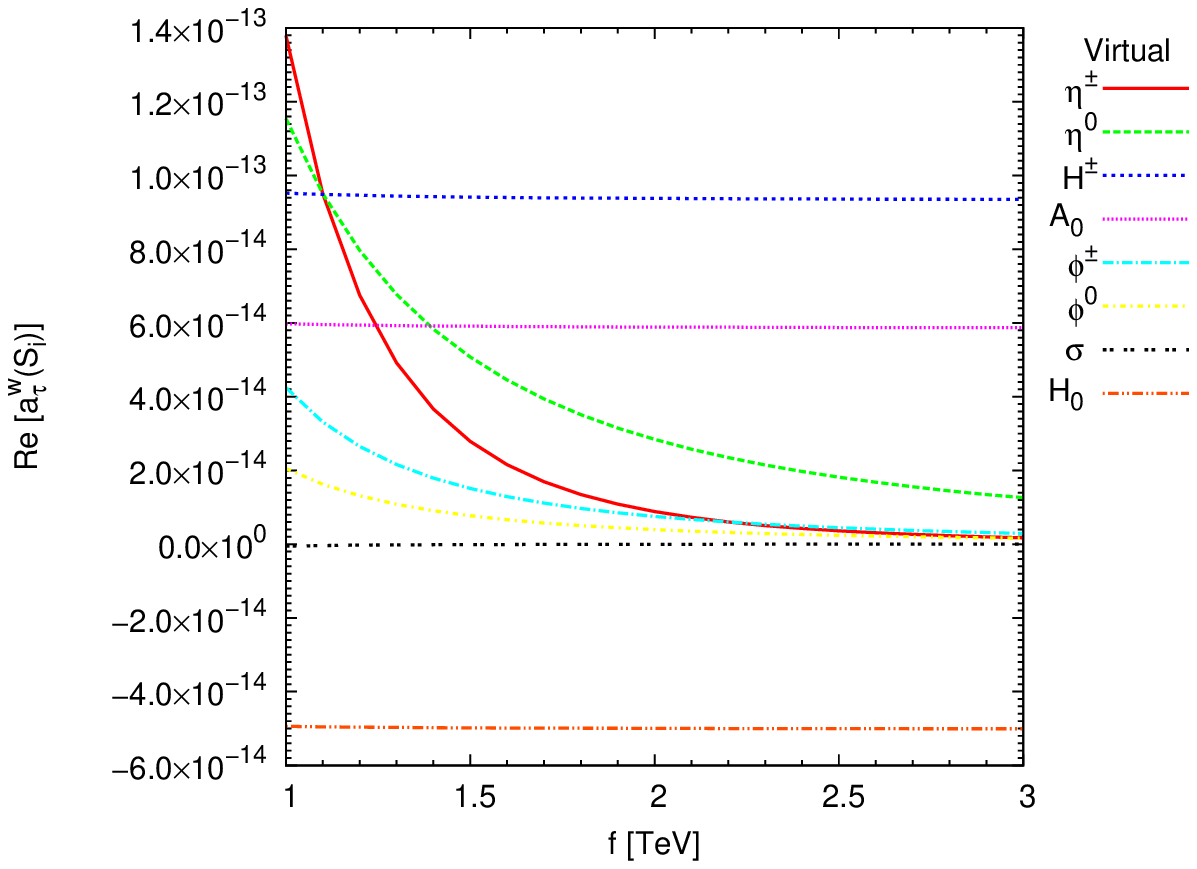}}
\subfloat[]{\includegraphics[width=8.0cm]{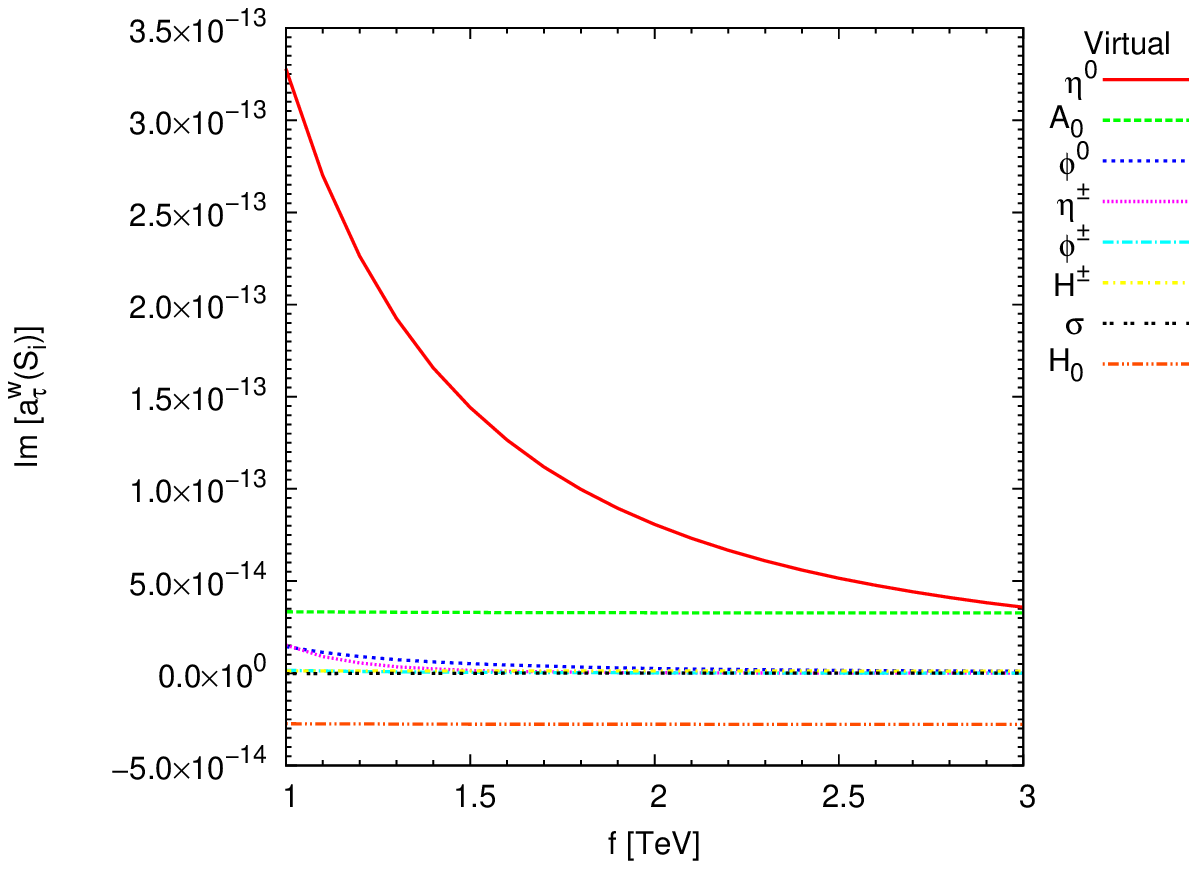}}
\caption{ \label{Si-weak} Individual scalar contributions to $ a^{W}_{\tau}$.
 a) Re($ a^{W}_{\tau}$). b) Im($ a^{W}_{\tau}$).
The plots are obtained with the fixed value of $F=4000$ GeV. For the remaining model parameters, the values provided in Table~\ref{parametervalues} are used. 
 }
\end{figure}

In this subsection, we perform the numerical estimation of the AWMDM $ a^{W}_{\tau} $ ($ a^{W}_{\tau }\equiv a^{W-BLHM}_{\tau} $) of the tau-lepton induced by scalar and vector bosons of the BLHM, these new particles that contribute to $ a^{W}_{\tau}$ are $A_0$, $ H_0 $, $ \phi^{0} $, $ \eta^{0} $, $ \sigma $, $ H^{\pm} $, $ \phi^{\pm} $, $ \eta^{\pm} $, $ W' $ and $ Z' $.
In this sense, we start by showing in Fig.~\ref{Si-weak} the contributions of the different scalars to the $\tau$ AWMDM, 
here and in the subsequent cases $ a^{W}_{\tau}$  acquire a real part and an imaginary part.

\noindent In this figure, we plot the behavior of
 $ a^{W}_{\tau}$ as a function of the new physical scale $f$ for the interval $f=[1000, 3000]$ GeV, while the other parameters assume fixed values.
 In the left plot of Fig.~\ref{Si-weak}, it can be seen that the great majority of the scalars involved generate positive contributions to  Re$\,[ a^{W}_{\tau}]$, and only the   $ H_{0}$ and $\sigma$ scalars contribute negatively. 
Of all the scalars, the heaviest of them is $ \sigma $ and it contributes quite small values to Re$\,[ a^{W}_{\tau}]$, $|\text{Re}\,[ a^{W}_{\tau}(\sigma)]|=[5.16\times 10^{-16}, 8.20\times 10^{-18}]$.
In contrast, the scalars $\eta^{\pm}$ and $H^{ \pm}$ generate the main contributions to Re$\,[ a^{W}_{\tau}]$  in the range of analysis established for the $f$ energy scale, i.e, Re$\,[a^{W}_{\tau}(\eta^{\pm} )]=[1.38\times 10^{-13}, 9.51 \times 10^{-14}]$ for the interval $f=[1000,1100]$ GeV and Re$\,[ a^{W}_{\tau}(H^{\pm} )]=(9.51, 9.35] \times 10^{-14}$  for $f=(1100,3000]$ GeV. 
With respect to the right plot of Fig.~\ref{Si-weak}, it can be observed that again  the $ H_0 $ and $ \sigma $ scalars contribute negatively, in this case to Im$\,( a^{W}_{\tau})$, while the rest of the scalars contribute positively. The $\eta^{0}$ scalar provides the largest contributions to Im$\,[ a^{W}_{\tau}]$, while the smallest contribution is induced by $ \sigma $: Im$\,[ a^{W}_{\tau}(\eta^{0})]=[3.28\times 10^{-13},3.58\times 10^{-14}]$ and $|\text{Im}\,[ a^{W}_{\tau}(\sigma)]|=[1.14 \times 10^{-16},1.38\times 10^{-18}]$.

\begin{figure}[H]
\center
\subfloat[]{\includegraphics[width=8.0cm]{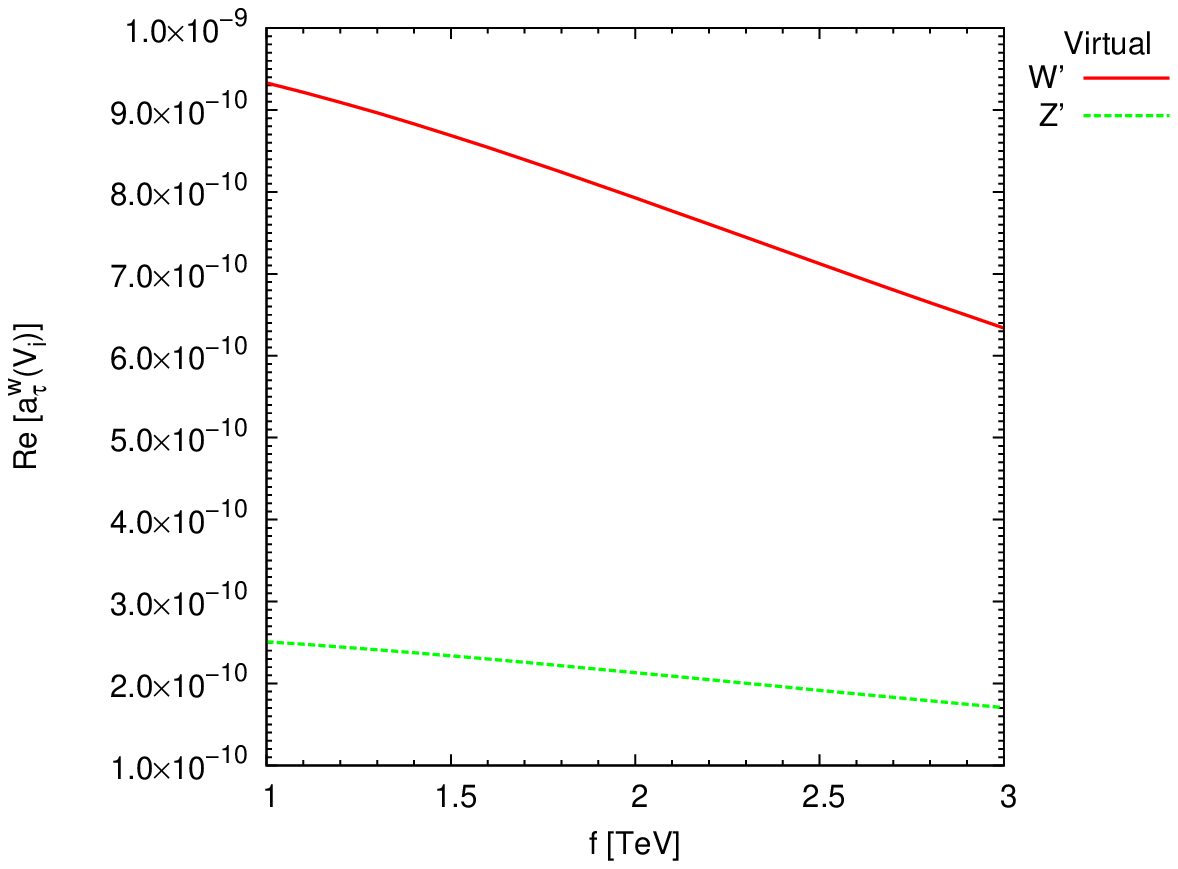}}
\subfloat[]{\includegraphics[width=8.0cm]{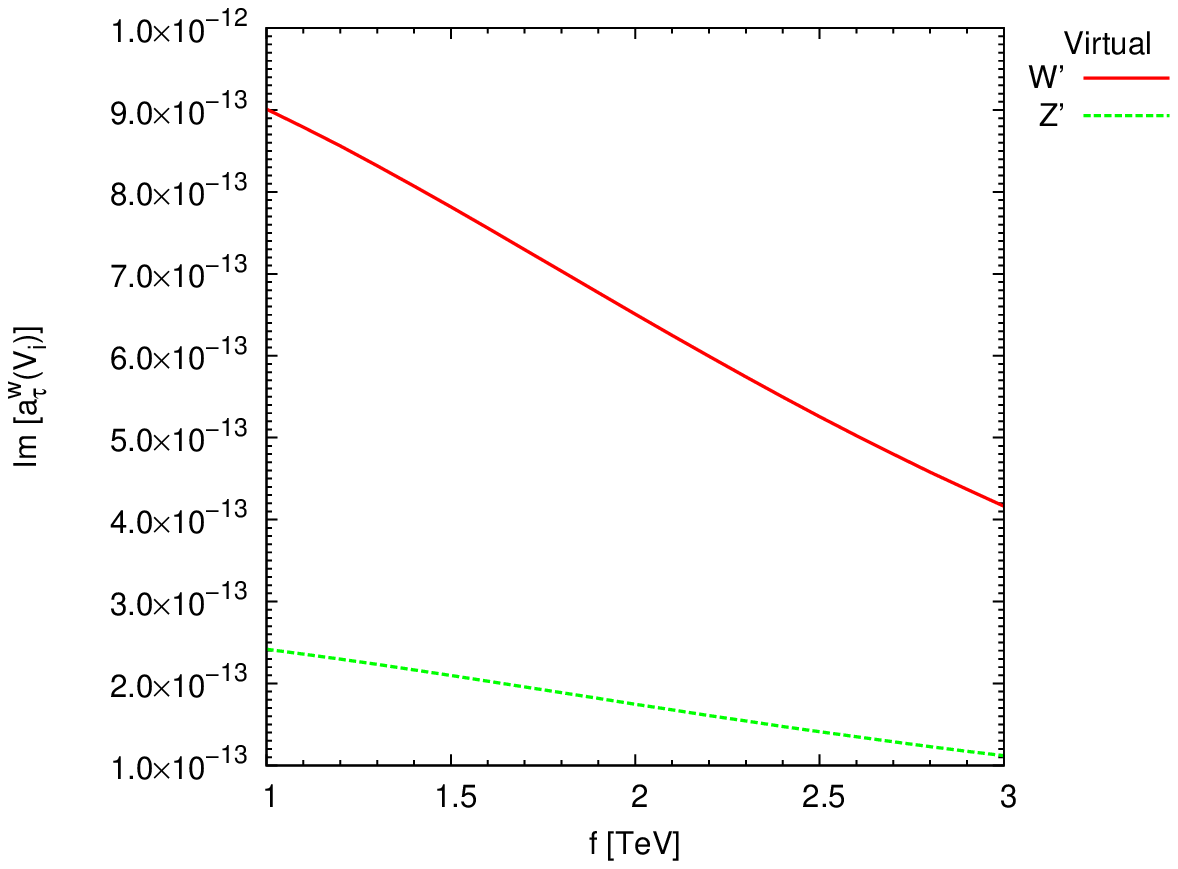}}
\caption{ \label{Vi-weak} Individual vector contributions to $ a^{W}_{\tau}$.
 a) Re($ a^{W}_{\tau}$). b) Im($ a^{W}_{\tau}$).
The plots are obtained with the fixed value of $F=4000$ GeV. For the remaining model parameters, the values provided in Table~\ref{parametervalues} are used. 
 }
\end{figure}

We discuss the contributions induced by the vector bosons, $W'$ and $Z'$, to the $\tau  $ AWMDM.  We begin by examining the real and imaginary parts of the BLHM partial contributions to $ a^{W}_{\tau}$. Thus, from Fig.~\ref{Vi-weak}, we can see that all the generated contributions are positive.  
 The highest signals for both the real and imaginary part of $ a^{W}_{\tau}$ are achieved when the vector boson $W'$ circulates in the $Z\tau^{+} \tau^{-}$ vertex loop. In this case, the corresponding numerical contributions are  Re$[ a^{W}_{\tau}(W')]=[9.33,634]\times 10^{-10}  $ and Im$[ a^{W}_{\tau}(W')]=[9.01,4.16]\times 10^{-13}$. 
Complementarily, the weakest signals appear when the gauge boson $Z'$ circulates in the above-mentioned vertex loop, these contributions are Re$[ a^{W}_{\tau}(Z')]=[2.51,1.70]\times 10^{-10}  $ and Im$[ a^{W}_{\tau}(Z')]=[2.42,1.12]\times 10^{-13}$.
  If we compare our numerical estimates, we find that the real parts of the partial contributions provide significant contributions to $ a^{W}_{\tau}$ as these are three orders of magnitude larger than the imaginary parts.
In Table~\ref{parcialweak} we show the magnitudes of all parcial contributions to $  a^{W}_{\tau} $ that correspond to the virtual particles circulating in the $Z\tau^{+} \tau^{-}$ vertex loop.

\begin{figure}[H]
\center
\subfloat[]{\includegraphics[width=8.0cm]{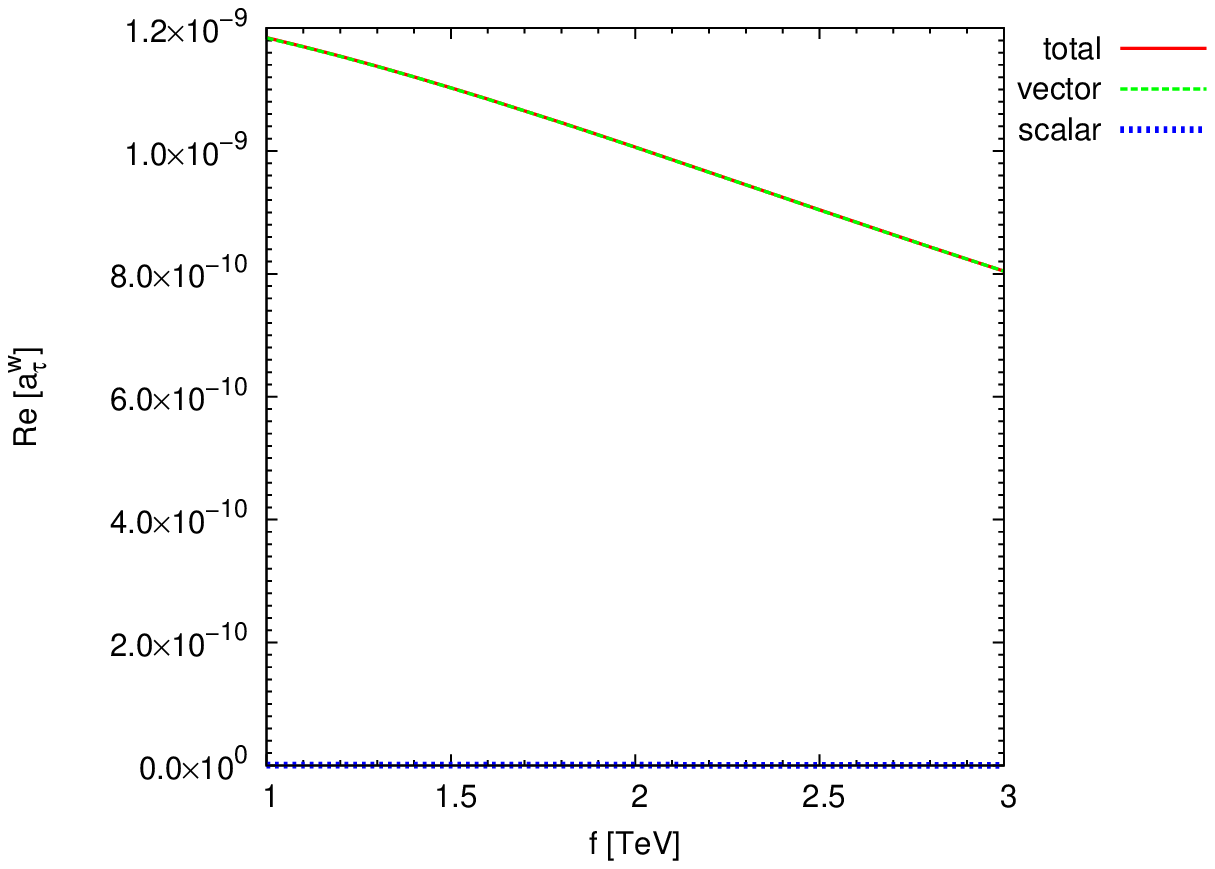}}
\subfloat[]{\includegraphics[width=8.0cm]{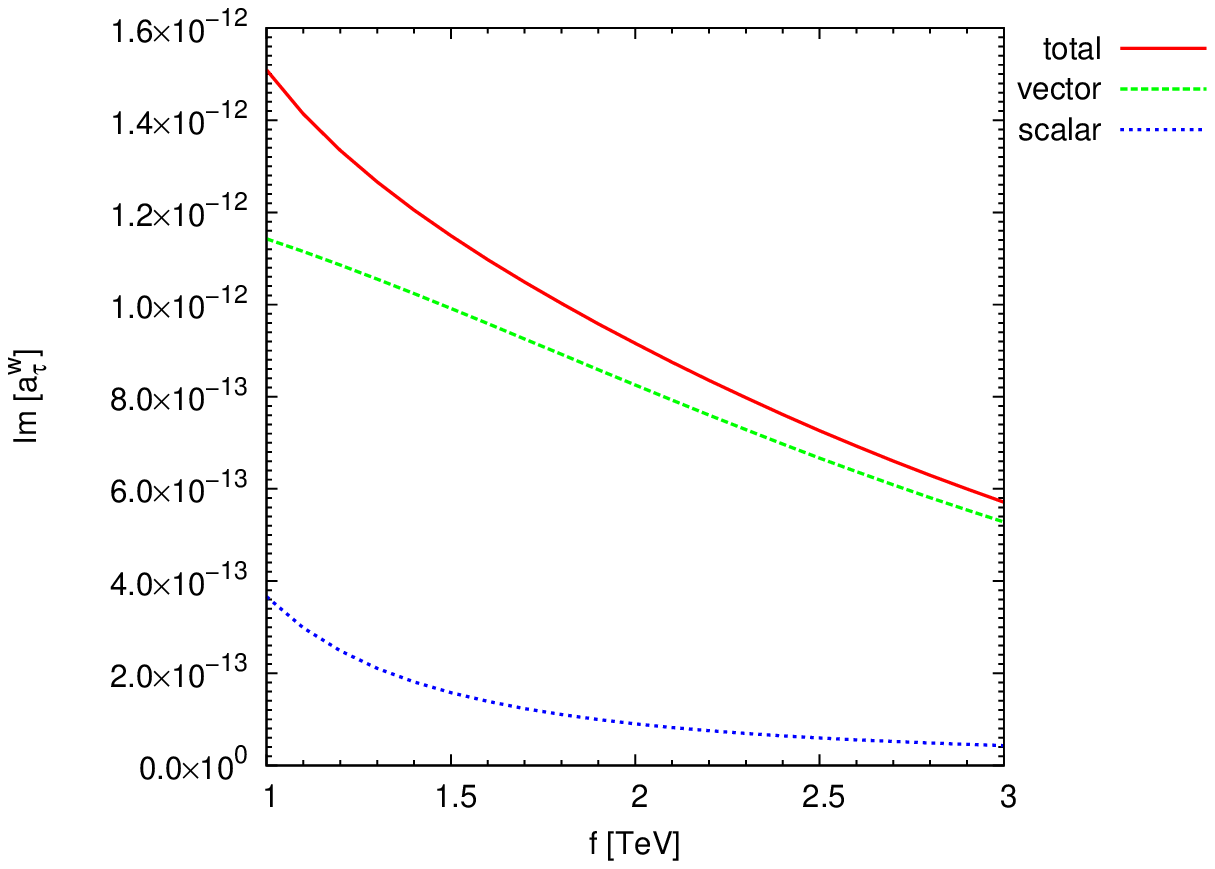}}
\caption{ \label{SVT-weak} Scalar, vector and total contributions to $ a^{W}_{\tau}$.
 a) Re($ a^{W}_{\tau}$). b) Im($ a^{W}_{\tau}$).
The plots are obtained with the fixed value of $F=4000$ GeV. For the remaining model parameters, the values provided in Table~\ref{parametervalues} are used. 
 }
\end{figure}

In the following, we show the curves that represent the sum of all individual contributions due to the scalar and vector bosons.
The magnitude of these contributions are shown in Fig.~\ref{SVT-weak}, here we observe that the total vector contribution dominates over the total scalar contribution since the latter is suppressed at least one order of magnitude more than the first one.
 With respect to the real part of $ a^{W}_{\tau}$ depicted in Fig.~\ref{SVT-weak}(a), we can observe more closely that Re$[ a^{W}_{\tau}(V)]  $ and Re$[ a^{W}_{\tau}(S+V)]  $ obtain values of the same order of magnitude, $ 10^{-9}-10^{-10} $, while Re$[ a^{W}_{\tau}(S)] \sim 10^{-13} $.
With the imaginary part of $ a^{W}_{\tau}$ (see Fig.~\ref{SVT-weak}(b)) the same happens as the real part, in this case, Im$[ a^{W}_{\tau}(V)]  $ $ \sim $ Im$[ a^{W}_{\tau}(S+V)]  $  $ \sim 10^{-12}-10^{-13} $ and Im$[ a^{W}_{\tau}(S)] \sim 10^{-13}-10^{-14} $.
 It is worth mentioning that Figs.~\ref{SVT-weak}(a) and~\ref{SVT-weak}(b) were obtained for a fixed value of the other physical scale of the BLHM, $F=4000$ GeV, which is also involved in our calculations. 
 We later present the sensitivity of $ a^{W}_{\tau}$ for other values of the  $F$ scale.

\begin{figure}[H]
\center
\subfloat[]{\includegraphics[width=8.0cm]{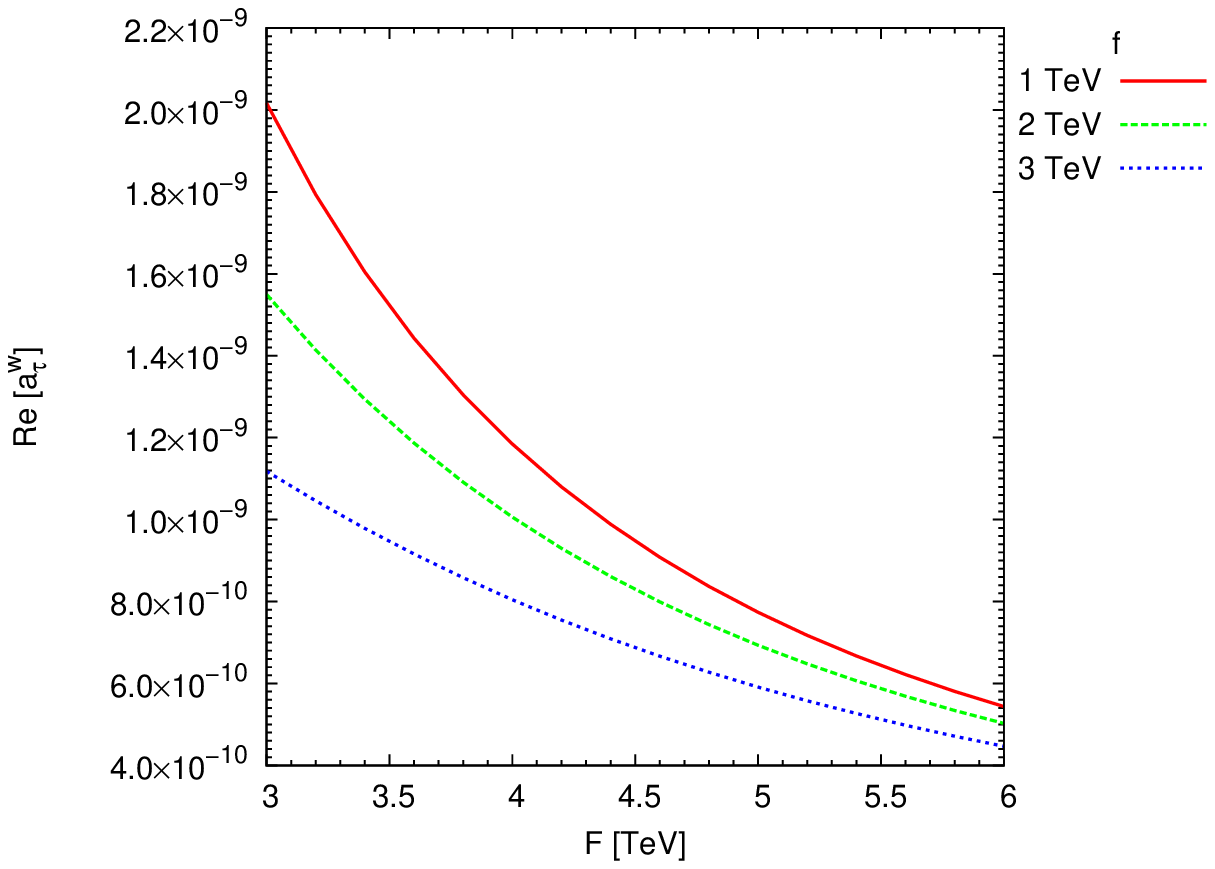}}
\subfloat[]{\includegraphics[width=8.0cm]{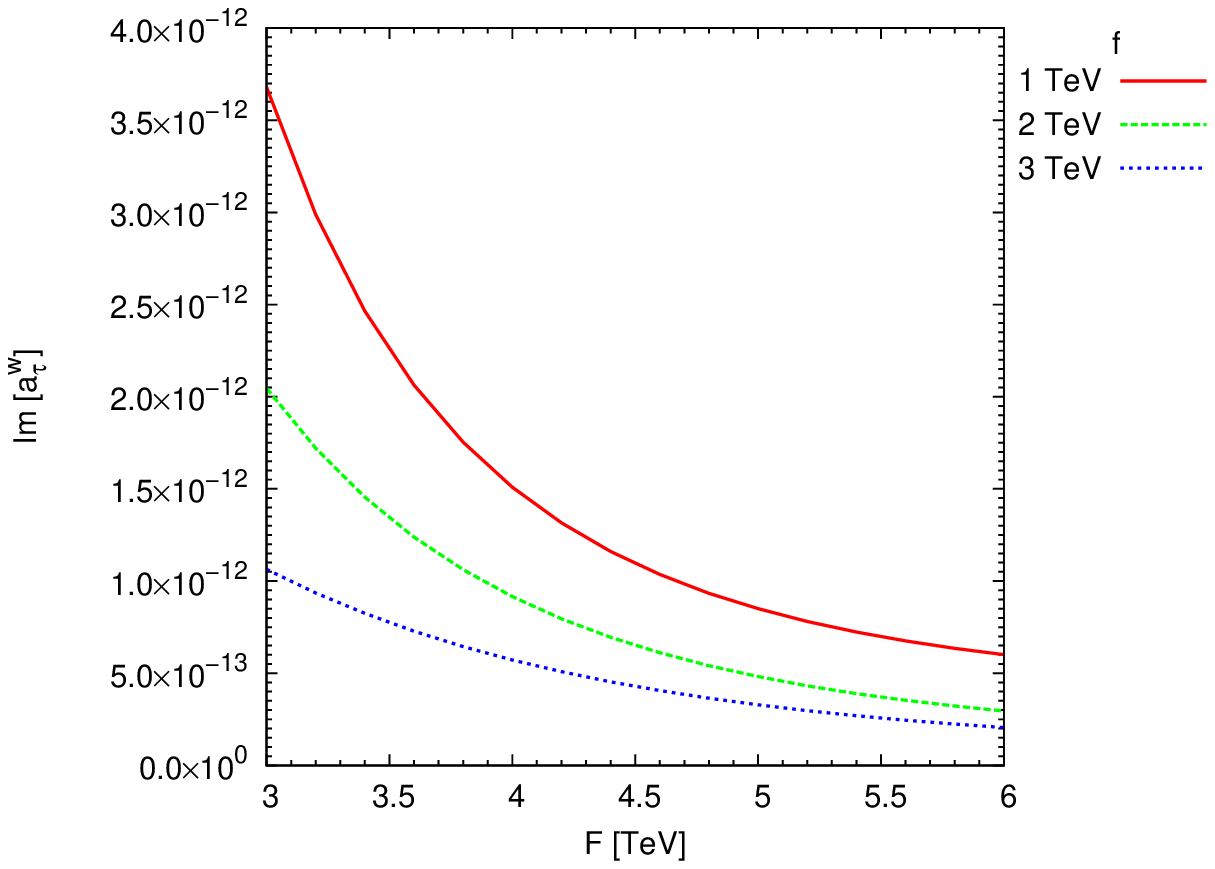}}
\caption{ \label{fi-weak} Total contribution to $ a^{W}_{\tau}$ for different values of the energy scale $f$.
 a) Re($a^{W}_{\tau}$). b) Im($ a^{W}_{\tau}$).
The plots are obtained for certain fixed values of the $f$ scale. For the remaining parameters of the model, the values provided in Table~\ref{parametervalues} are used. 
 }
\end{figure}

As already commented, the  BLHM is based on two distinct global symmetries that break to diagonal subgroups at different scales, $f$ and $F>f$, these scales represent the scales of the new physics. 
Therefore, it is very convenient to analyze the behavior of the $\tau$ AWMDM as a function of these energy scales, since the masses of the new scalar and vector bosons strongly depend on them.
Thus, similar to what was performed in the previous subsection, we  study the dependence of $ a^{W}_{\tau}$ on $F$ while maintaining fixed the $f$ scale or the opposite.
In Fig.~\ref{fi-weak} we begin by showing a variation of the scale $F$ from 3000 GeV to 6000 GeV, for three different $f$ energy scales, i.e., $f=1000$ GeV, $2000$ GeV, $3000$ GeV.
In this plot we appreciate that the main contributions to $ a^{W}_{\tau}$ arise for $f=1000$ GeV, this occurs for both the real and the imaginary part of $ a^{W}_{\tau}$:  Re$[ a^{W}_{\tau}]=[2.02\times 10^{-9},5.44\times 10^{-10}] $ and  Im$[ a^{W}_{\tau}]=[3.68\times 10^{-12},6.00\times 10^{-13}]$, respectively.
On the other hand, the weakest contributions appear when the scale $f$ takes larger values, especially when $f=3000$ GeV: Re$[ a^{W}_{\tau}]=[1.12\times 10^{-9},4.47\times 10^{-10}] $ and  Im$[ a^{W}_{\tau}]=[1.06\times 10^{-12},2.05\times 10^{-13}]$.

\begin{figure}[H]
\center
\subfloat[]{\includegraphics[width=8.0cm]{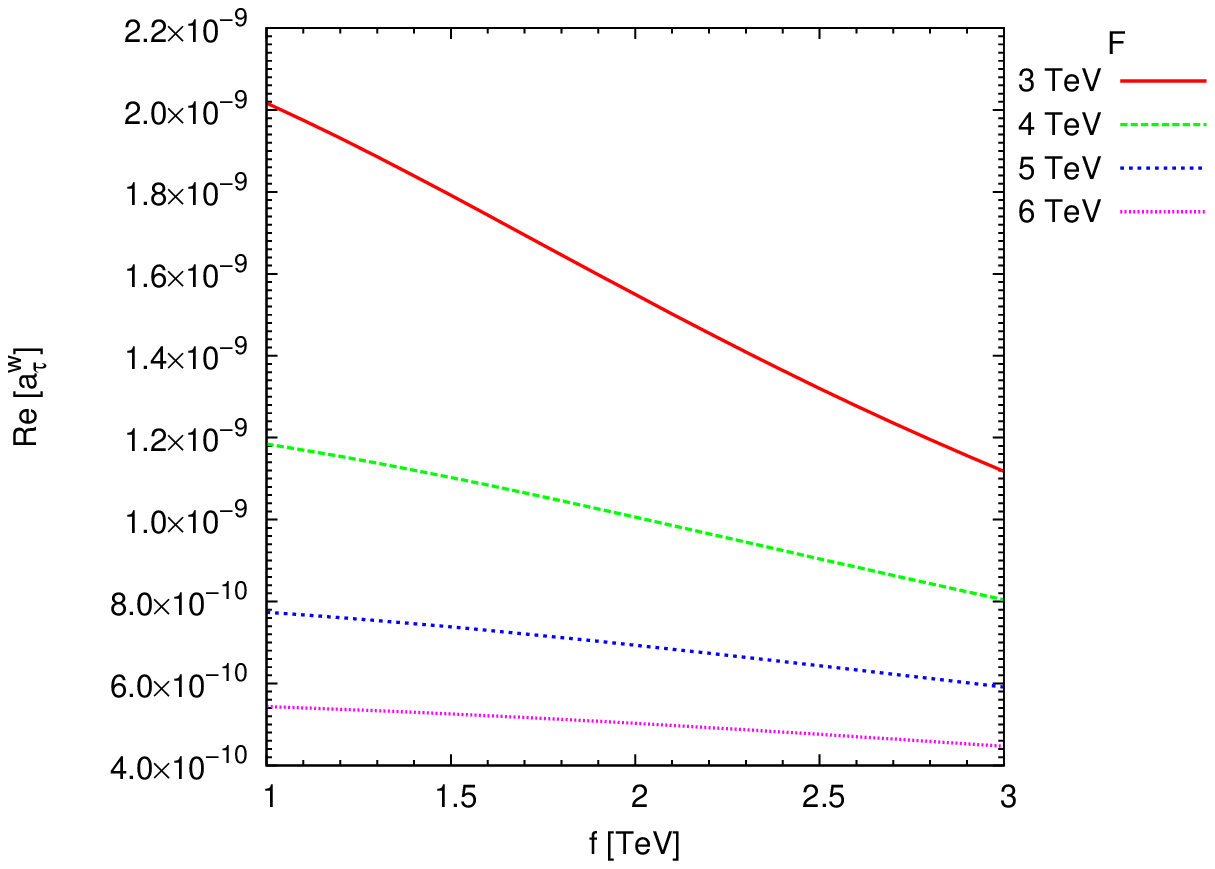}}
\subfloat[]{\includegraphics[width=8.0cm]{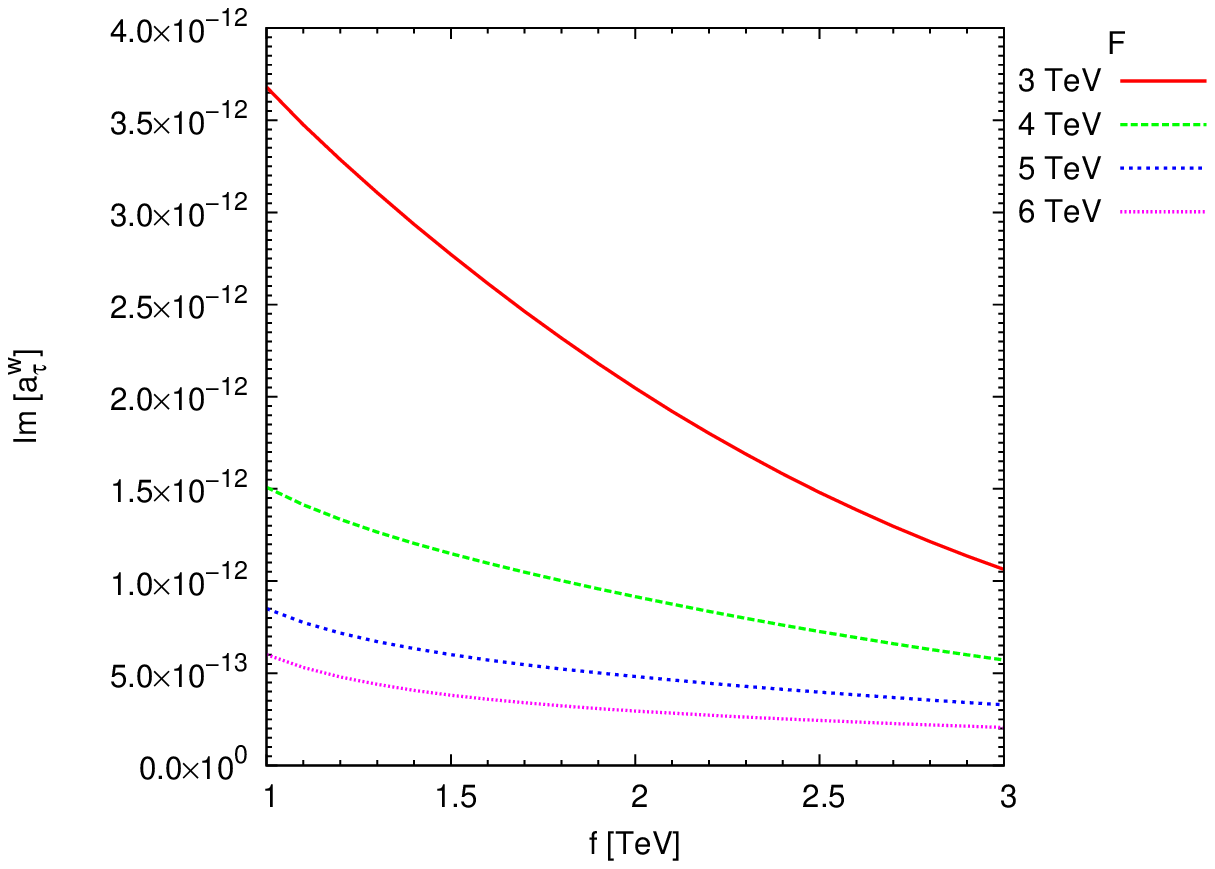}}
\caption{ \label{Fi-weak} Total contribution to $a^{W}_{\tau}$ for different values of the energy scale $F$. a) Re($ a^{W}_{\tau}$). b) Im($a^{W}_{\tau}$).
The plots are obtained for certain fixed values of the $F$ scale. For the remaining parameters of the model, the values provided in Table~\ref{parametervalues} are used. 
}
\end{figure}

Now, we examine the dependence of $ a^{W}_{\tau}$ on the $f$ scale for certain fixed values of the $F$ scale, i.e., $ F=3000 $ GeV, $ 4000 $ GeV, $ 5000 $ GeV, $ 6000 $ GeV. With these values of $F$ we plot the curves shown in Fig.~\ref{Fi-weak}. 
 In this case, the largest contributions to $ a^{W}_{\tau}$ are reached for $F=3000$ GeV, these contributions are Re$[ a^{W}_{\tau}]=[2.02, 1.12]\times 10^{-9} $ and  Im$[ a^{W}_{\tau}]=[3.68,1.06]\times 10^{-12}$.
On the opposite side, the smallest contributions to $ a^{W}_{\tau}$  are generated when $F=6000$ GeV: Re$[ a^{W}_{\tau}]=[5.44, 4.47]\times 10^{-10} $ and  Im$[ a^{W}_{\tau}]=[6.00,2.05]\times 10^{-13}$.

According to the numerical results,  it is found that $ a^{W}_{\tau}$ is sensitive to a slight change in the values of the $f$ and $F$ scales, this occurs as long as these parameters are in the established intervals. 
When  $ a^{W}_{\tau}$ depends on $F$, we observe that  $ a^{W}_{\tau}$ has a decrease of about one order of magnitude as $F$ increases up to 6000 GeV. For the next case, when  $ a^{W}_{\tau}$ depends on $f$, we also have a decrease of at most one order of magnitude as $f$ reaches 3000 GeV.
In summary, we can affirm that $ a^{W}_{\tau}$ gets large values when  $f=1000$ GeV or $F=3000$ GeV, while smaller values are obtained for $ a^{W}_{\tau}$  when the scales tend to take values close to their established upper limits.
In short, $ a^{W}_{\tau}=2.02\times 10^{-9} + 3.68\times 10^{-12} \,i$ is the largest value found when $f=1000$ GeV and $F=3000$ GeV.

\begin{figure}[H]
\center
\subfloat[]{\includegraphics[width=8.2cm]{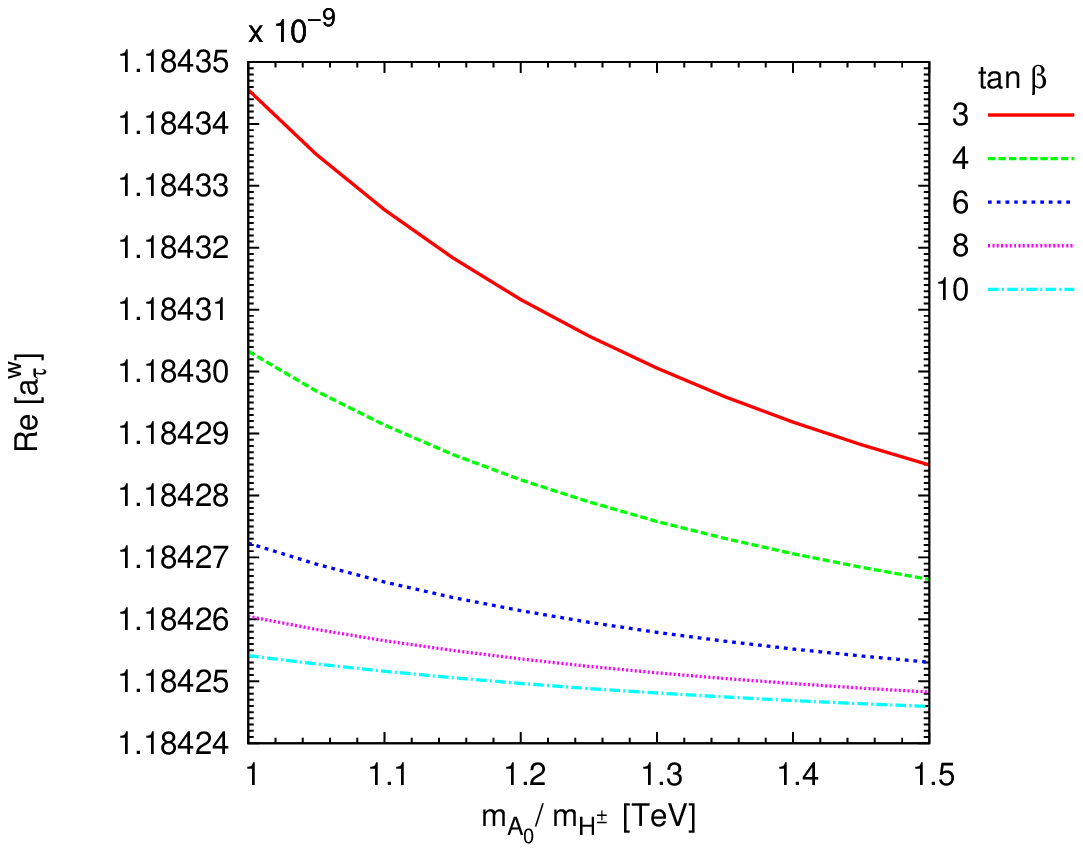}}
\subfloat[]{\includegraphics[width=8.2cm]{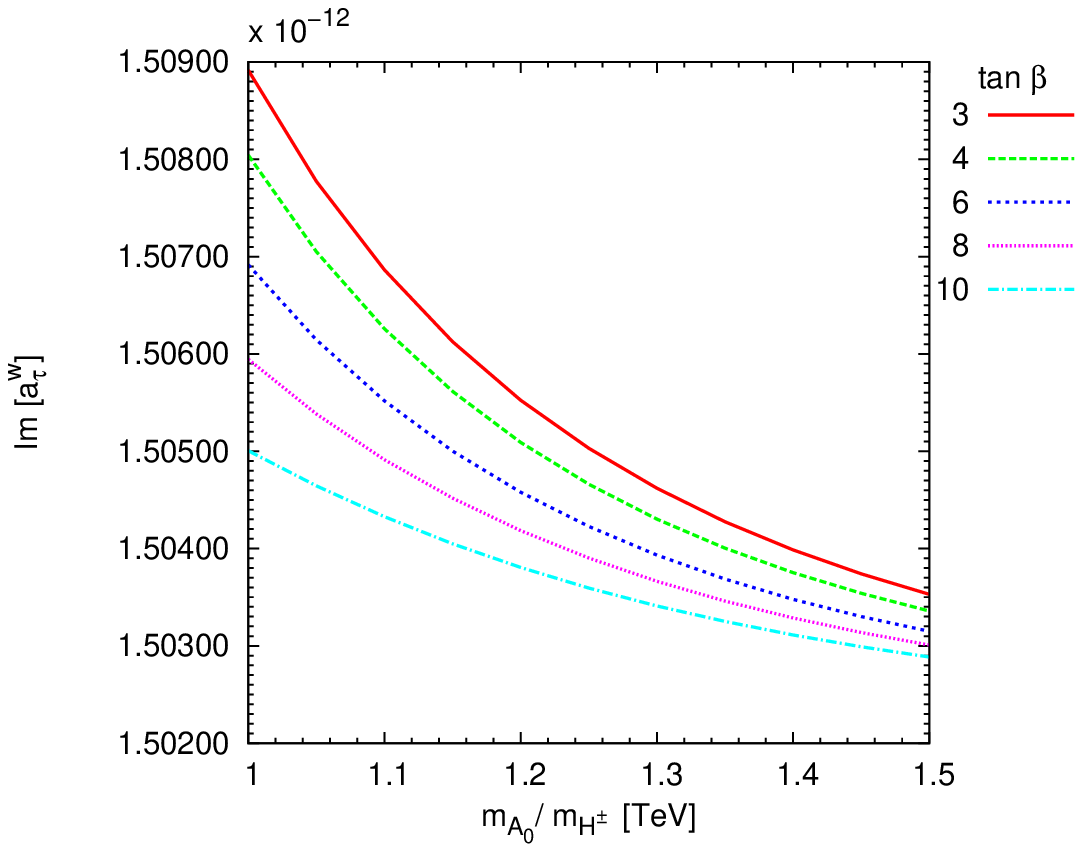}}
\caption{ \label{beta-weak} Total contribution to $ a^{W}_{\tau}$ for different values of $\tan \beta$. a) Re($ a^{W}_{\tau}$). b) Im($ a^{W}_{\tau}$). 
The plots are obtained for fixed values of the $f=1000$ GeV and  $F=4000$ GeV. For the remaining parameters of the model, the values provided in Table~\ref{parametervalues} are used.}
\end{figure}

Finally, Fig.~\ref{beta-weak} show the $ a^{W}_{\tau}$ behavior as a function of $ m_{A_0} $ or $ m_{H^{\pm}} $, for $\tan \beta=3$, $ 4 $, $ 6 $, $ 8 $ and $ 10 $.
From this figure, we can observe that the largest contributions to the $ \tau $ AWMDM arise when $\tan \beta =3$, this happens for the real and imaginary part of $ a^{W}_{\tau}$: Re$[ a^{W}_{\tau}]=[1.18435, 1.18428]\times 10^{-9} $ and  Im$[ a^{W}_{\tau}]=[1.50892,1.50353]\times 10^{-12}$.
As the parameter $\tan \beta$ increases, more suppressed curves are generated, this occurs in the case of $\tan \beta=10$ which gives the following values corresponding to $ a^{W}_{\tau}$,  Re$[ a^{W}_{\tau}]=[1.18425, 1.18424]\times 10^{-9} $ and  Im$[ a^{W}_{\tau}]=[1.50500,1.50288]\times 10^{-12}$.
Furthermore, $ a^{W}_{\tau}$ acquires smaller values as the mass of the pseudoscalar $ A_0 $ increases up to 1500 GeV. However, we can say that the changes or effects on $ a^{W}_{\tau}$ are not so great, since the numerical values they acquire remain of the same order of magnitude.

\begin{table}[H]
\caption{The magnitude of the partial contributions to $a^{W}_{\tau}$ of the BLHM.
The data are obtained by fixing the $f$ and $F$ scales, $f=1000$ GeV and $F=4000$ GeV. For the rest of the model parameters, the values provided in Table~\ref{parametervalues} are used. {\bf abc} denotes the different particles running in the loop of the vertex $Z \tau^{+} \tau^{-}$.
\label{parcialweak}}
\centering
\begin{tabular}{|c|c|}
\hline
\hline
\multicolumn{2}{|c|}{ $f=1 000\ \text{GeV}$, $F=4000\ \text{GeV}$}\\
\hline
 $\text{Couplings}\  \textbf{abc}$  & $\left(a^{W}_\tau \right)^{ \textbf{abc} } $  \\
\hline
\hline
$\sigma \tau \tau $  & $- 2_\cdot 83\times 10^{-16} - 1_\cdot 14 \times 10^{-16}\ i$  \\
\hline
$A_{0} \tau \tau  $  & $ 5_\cdot 98 \times 10^{-14} +  3_\cdot 33\times 10^{-14}\ i$  \\
\hline
$H_{0} \tau \tau  $  & $ -4_\cdot 94\times 10^{-14} - 2_\cdot 74\times 10^{-14}\ i$  \\
\hline
$\eta^{0} \tau \tau  $  & $ 1_\cdot 15 \times 10^{-13} + 3_\cdot 28 \times 10^{-13}\ i$  \\
\hline
$\phi^{0} \tau \tau  $  & $ 2_\cdot 05 \times 10^{-14} + 1_\cdot 44 \times 10^{-14}\ i$  \\
\hline
$H^{\pm} \nu_{\tau} \nu_{\tau}  $  & $ 9_\cdot 52\times 10^{-14} + 1_\cdot 21\times 10^{-15}\ i$  \\
\hline
$\eta^{\pm} \nu_{\tau} \nu_{\tau}  $  & $ 1_\cdot 38 \times 10^{-13} + 1_\cdot 53 \times 10^{-14}\ i$  \\
\hline
$\phi^{\pm} \nu_{\tau} \nu_{\tau}  $  & $ 4_\cdot 25 \times 10^{-14} + 1_\cdot 65 \times 10^{-15}\ i$  \\
\hline
$Z' \tau \tau  $  & $ 2_\cdot 51 \times 10^{-10} + 2_\cdot 42 \times 10^{-13}\ i$  \\
\hline
$W'^{\pm} \nu_{\tau} \nu_{\tau}  $  & $ 9_\cdot 33 \times 10^{-10} + 9_\cdot 01 \times 10^{-13}\ i$  \\
\hline
\end{tabular}
\end{table}

\subsection{Contributions of the SM and BLHM to the AMDM and AWMDM of the tau-lepton}

As previously defined (Eqs.~(\ref{atau-total}) and~(\ref{awtau-total})), the contribution of the SM and BLHM particles to the AMDM and AWMDM of the tau-lepton will be represented by $ \alpha_\tau $ and $ \alpha^{W}_{\tau} $, respectively. In this way, in Tables~\ref{suma-atau} and~\ref{suma-awtau} we provide the partial and total numerical values for  $ \alpha_\tau $ and $ \alpha^{W}_{\tau} $. In these tables we find that all new diagrams arising in the BLHM have a small numerical impact on  the AMDM and AWMDM of the $ \tau $-lepton. 
This is partly because both $ a_\tau $ and $ a^{W}_{\tau} $ acquire values inversely proportional to the energy scales $f$ and $F$. Only some of the partial contributions are comparable to the SM partial contributions (see Appendix B), but not larger than them. 
The SM particles provide the largest contributions to $ \alpha_\tau $ and $ \alpha^{W}_{\tau} $. According to the numerical data, we find that $ \alpha_\tau = 116090_\cdot 37\times 10^{-8} $ and $ \alpha^{W}_{\tau} = -1_\cdot 92\times 10^{-6} - 0_\cdot 57\times 10^{-6}\ i $.

\begin{table}[H]
\caption{Partial contributions to $ \alpha_\tau $. {\bf abc} denotes the different particles running in the loop of the vertex $\gamma \tau^{+} \tau^{-}$.
\label{suma-atau}}
\centering
\begin{tabular}{|c|c|}
\hline
\hline
\multicolumn{2}{|c|}{ $f=1 000\ \text{GeV}$, $F=4000\ \text{GeV}$}\\
\hline
 $\text{Couplings}\  \textbf{abc}$  & $\left(\alpha_\tau \right)^{ \textbf{abc} } $  \\
\hline
$ \gamma \tau \tau   $  & $ 116140_\cdot  97 \times 10^{-8}+ 0 \,i $  \\
\hline
$ Z \tau \tau + Z'\tau \tau$  & $ 51_\cdot 59 \times 10^{-8} + 0 \ i$  \\
\hline
$h_0\tau \tau + A_0 \tau \tau + H_0 \tau \tau +\phi^{0}\tau \tau + \eta^{0} \tau \tau + \sigma \tau \tau $  & $ 0.09 \times 10^{-8} +0 \ i$  \\
\hline
$  \nu_\tau WW $  & $ -102_\cdot 29\times 10^{-8}+ 0 \ i$  \\
\hline
$ \textbf{Total} $  & ${\bf \alpha_\tau }= 116090_\cdot 37\times 10^{-8} +0\ i$  \\
\hline
\end{tabular}
\end{table}

\begin{table}[H]
\caption{Partial contributions to $ \alpha^{W}_\tau $. {\bf abc} denotes the different particles running in the loop of the vertex $Z \tau^{+} \tau^{-}$.
\label{suma-awtau}}
\centering
\begin{tabular}{|c|c|}
\hline
\hline
\multicolumn{2}{|c|}{ $f=1 000\ \text{GeV}$, $F=4000\ \text{GeV}$}\\
\hline
 $\text{Couplings}\  \textbf{abc}$  & $\left(\alpha^{W}_\tau \right)^{ \textbf{abc} } $  \\
\hline
$ \gamma \tau \tau  $  & $ 3_\cdot 09 \times 10^{-7}-1_\cdot 24 \times 10^{-7} \,i $  \\
\hline
$ Z \tau \tau + Z'\tau \tau  $  & $ 4_\cdot 05 \times 10^{-8} + 1_\cdot 86\times 10^{-8} \ i$  \\
\hline
$h_0\tau \tau  + A_0 \tau \tau + H_0 \tau \tau +\phi^{0}\tau \tau + \eta^{0} \tau \tau + \sigma \tau \tau  $  & $ -6_\cdot 58\times 10^{-12} - 1_\cdot 34\times 10^{-11} \ i$  \\
\hline
$ W\nu_\tau \nu_\tau + W' \nu_\tau \nu_\tau + H^{\pm}\nu_\tau \nu_\tau + \phi^{\pm}\nu_\tau \nu_\tau + \eta^{\pm}\nu_\tau \nu_\tau$  & $ -9_\cdot 13\times 10^{-7} -4_\cdot 66\times 10^{-7} \ i$  \\
\hline
$  \nu_\tau WW$  & $ -1_\cdot 37\times 10^{-6}+ 0 \ i$  \\
\hline
$ \tau h_0 Z  $  & $ 1_\cdot 35 \times 10^{-8} +0 \ i$  \\
\hline
$ \textbf{Total} $  & ${\bf \alpha^{W}_{\tau} }= -1_\cdot 92\times 10^{-6} - 0_\cdot 57\times 10^{-6}\ i$  \\
\hline
\end{tabular}
\end{table}

\section{Conclusions}

We have calculated at the one-loop level the contributions generated by the SM (see Appendix B) and BLHM particles to the AMDM and AWMDM of the tau-lepton. Within the SM, we find that our predictions for $ a^{SM}_{\tau} $ and $ a^{W-SM}_{\tau} $ are in agreement with results reported in the literature. With respect to the new physics, this arises in the BLHM scenario and are induced by the new scalar and vector bosons of the model. The new contributions that these generate to $ a_{\tau} $ and $ a^{W}_{\tau} $ are emphasized.

The BLHM has the characteristic of having two different global symmetries that break at different energy scales, so $f$ and $F$ represent the scales of the new physics, and at this level the new scalar and vector bosons acquire their respetive masses. 
Therefore, we have analyzed the dependence of $a_{\tau} $ and $ a^{W}_{\tau} $ on the physical scales $ f $ and $ F $,  and we find that both $a_{\tau} $ and $ a^{W}_{\tau} $ are sensitive to changes in $ f $ and $ F $.
Large values of these energy scales, as long as they are in the allowed intervals, suppress the contributions to $a_{\tau} $ and $ a^{W}_{\tau} $. However, when these scales acquire the respective minimum values, $f=1000$ GeV and $F=3000$ GeV, large values are reached for the $ \tau $ AMDM and AWMDM: $a_{\tau} =6.92\times 10^{-10} + 0\, i$ and $ a^{W}_{\tau} =2.02\times 10^{-9} + 3.68\times 10^{-12}\, i$, respectively.
In this work, we also examine the dependence of $a_{\tau} $ and $ a^{W}_{\tau} $ on the $m_{A_0}$ parameter, our results indicate that both show a small sensitivity to changes in the $m_{A_0}$ parameter since the contributions they acquire remain of the same order of magnitude, $a_{\tau} \sim 10^{-10} $ and $ a^{W}_{\tau} \sim 10^{-9}$.

It is interesting to study the contributions of the new physics as they could provide a significant improvement in the AMDM and AWMDM of the tau-lepton. The reason is that, for now, the SM predictions on $a^{SM}_{\tau} $ and $ a^{SM-W}_{\tau} $ show 
a clear discrepancy with experimental measurements.  This discrepancy could be attributed to additional contributions with origin in new physics, i.e., such discrepancy could be attributed to the effect of new particles not described by the SM. Therefore, extensions of the SM are worth studying as they could generate large contributions of new physics that are close to the experimental limits.
In the BLHM scenario, we  found that the contributions generated by the new scalar and vector bosons to the $ \tau $ AMDM and AWMDM  are $a_{\tau} \sim 10^{-10} $ and $ a^{W}_{\tau} \sim 10^{-9}$. These numerical values are smaller than the SM contributions, however, they are similar in size and even larger than those arising in some SM extensions such as: 
the Simplest Little Higgs model, $ a_{\tau}\sim 10^{-9}$ and $ a^{W}_{\tau}\sim 10^{-9} $~\cite{Arroyo-Urena:2016ygo};
the Two-Higgs Doublet models (type-I and type-II), $ a^{W}_{\tau}\sim 10^{-9}-10^{-10} $~\cite{Arroyo-Urena:2018ygo,Bernabeu:1995gs};
the Two-Higgs Doublet models type-III with textures,  $ a_{\tau}\sim 10^{-7}-10^{-8}$  and $ a^{W}_{\tau}\sim 10^{-7}-10^{-10} $~\cite{Arroyo-Urena:2015uoa};
Scalar Leptoquark models,  $ a_{\tau}\sim 10^{-9}$ and $ a^{W}_{\tau}\sim 10^{-9} $~\cite{Bolanos:2013tda}; 
the Minimal Supersymmetric Standard model with a mirror fourth generation, $ a_{\tau}\sim 10^{-6}-10^{-10}$~\cite{Ibrahim:2008gg}; 
unparticle physics (for $ \bigwedge_{U}=10 $ TeV),  $ a_{\tau}\sim 10^{-9}-10^{-10}$ and $ a^{W}_{\tau}\sim 10^{-9}-10^{-10}$~\cite{Moyotl:2012zz}; the type-I and type-III seesaw models, $ |a^{I}_\tau|<1.87\times 10^{-8} $ and  $ |a^{III}_\tau|<7.55\times 10^{-9} $~\cite{Biggio:2008in}; and finally, in the framework of the effective lagrangian approach and the Fritzsch-Xing lepton mass matrix, $ a_{\tau}\sim 10^{-11}$~\cite{Huang:1999vb}.

Currently, the experimental limits in the $ \tau $ AMDM and AWMDM are well above theoretical predictions. Our results are also outside the detection range of future experiments so there is not yet sufficient sensitivity to be tested and cross-checked with experimental values.
The results presented here complement other studies performed on models with an extended scalar sector, and may be useful to the scientific community.

\vspace{7.0cm}

\begin{center}
{\bf Acknowledgements}
\end{center}

E. C. A. appreciates the post-doctoral stay. A. G. R. thank SNI and PROFEXCE (M\'exico).

\vspace{3cm}


\appendix

\section{The Feynman rules for the BLHM}

In this appendix we present the Feynman rules for the BLHM involved in our calculation for the AMDM and AWMDM of the tau-lepton.
It is convenient to define the following useful notation:

\begin{eqnarray}
c_{\beta} &=& \cos \beta, \\
 s_{\beta} &=& \sin \beta, \\
s_{\alpha} &=& \sin \alpha, \\
c_{\alpha} &=& \cos \alpha.
\end{eqnarray}

\begin{eqnarray}
y_{\tau }&=& \frac{m_\tau}{v \sin \beta} \left(1-\frac{v^{2}}{3 f^{2}} \right)^{-1/2}.
\end{eqnarray}

\begin{table}[H]
\caption{Feynman rules for the BLHM involving  the escalars  $\sigma$, $h_0$, $H_0$, $\phi^{0}$, $\eta^{0}$,  $H^{\pm}$, $\phi^{\pm}$, $\eta^{\pm}$, the pseudoscalar $ A_0 $ and the vector bosons $Z'$ and $W'$.
\label{FeyRul}}
\begin{tabular}{|c|p{14.3cm}|}
\hline
\textbf{Vertex} & \textbf{Feynman rules} \\
\hline
\hline
$ \sigma \bar{\tau} \tau $  & $   \frac{c_{\beta}  v y_{\tau}}{\sqrt{2} f}  $  \\
\hline
$ h_0 \bar{\tau} \tau $  & $  c_{\alpha} y_{\tau}-\frac{\left(16 c_{\beta}
   s_{\alpha} s_{\beta} v^2+c_{\alpha} \left(8
   c_{\beta}^2 v^2+24 s_{\beta}^2 v^2\right)\right)
   y_{\tau}}{24 f^2}   $  \\
\hline
$ H_0 \bar{\tau} \tau $  & $ -s_{\alpha} y_{\tau} +  \frac{\left(s_{\alpha} c_{\beta}^2-2 c_{\alpha}
   s_{\beta} c_{\beta}+3 s_{\alpha} s_{\beta}^2\right) v^2 y_{\tau}}{3 f^2}  $  \\
\hline
$ A_0 \bar{\tau} \tau $  & $   -i c_{\beta} y_{\tau}  $  \\
\hline
$ \eta^{0} \bar{\tau} \tau $  & $   -\frac{i s_{\beta} v y_{\tau}}{2 f}  $  \\
\hline
$ \phi^{0} \bar{\tau} \tau $  & $   \frac{i s_{\beta} v y_{\tau}}{2 f}  $  \\
\hline
$ H^{+} \bar{\nu}_\tau \tau $  & $   \sqrt{2} c_{\beta} y_{\tau}  P_{R} $  \\
\hline
$ \eta^{+} \bar{\nu}_\tau \tau $  & $  \frac{i s_{\beta} v y_{\tau}}{\sqrt{2} f}P_{R}   $  \\
\hline
$ \phi^{+} \bar{\nu}_\tau \tau $  & $  -\frac{i s_{\beta} v y_{\tau}}{\sqrt{2} f}P_{R}  $  \\
\hline
$Z' \bar{\tau} \tau$  & $ \frac{1}{2} i g \gamma^{\mu} P_{L}   $ \\
\hline
$W'^{+} \bar{\nu}_\tau \tau$  & $ - \frac{ig}{\sqrt{2}}  \gamma^{\mu} P_{L}   $ \\
\hline
\hline
\end{tabular}
\end{table}

\section{The  AMDM and AWMDM of the tau-lepton at the SM}

\begin{figure}[H]
\center
\subfloat[]{\includegraphics[width=4.5cm]{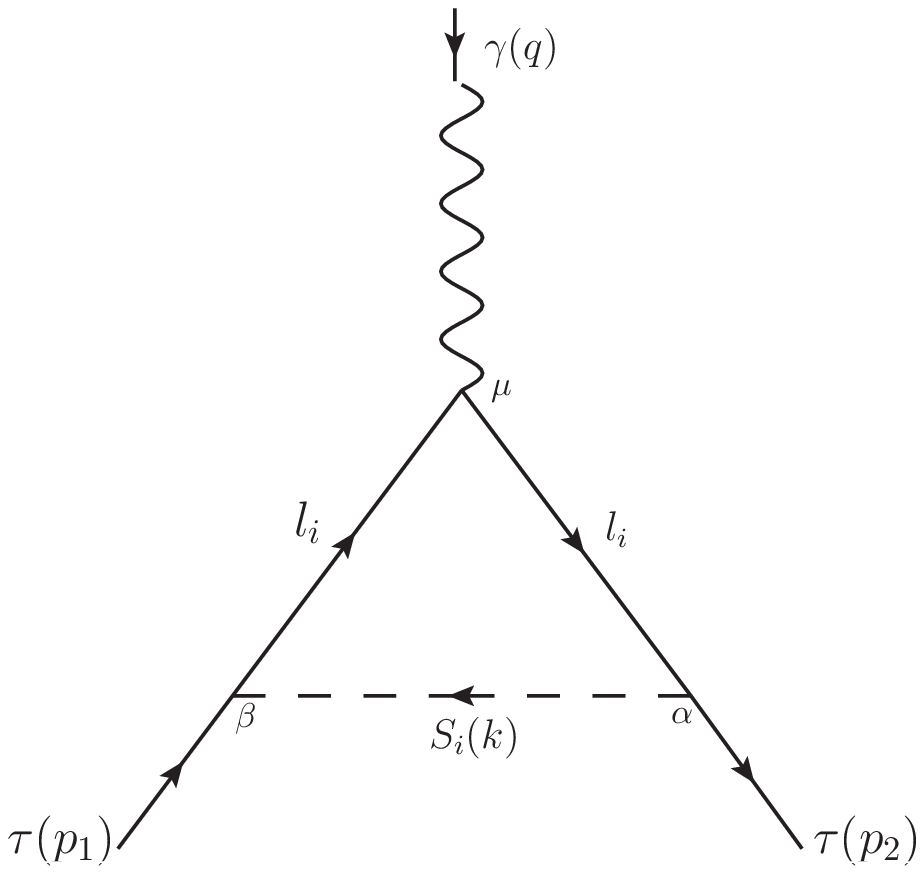}}
\subfloat[]{\includegraphics[width=4.5cm]{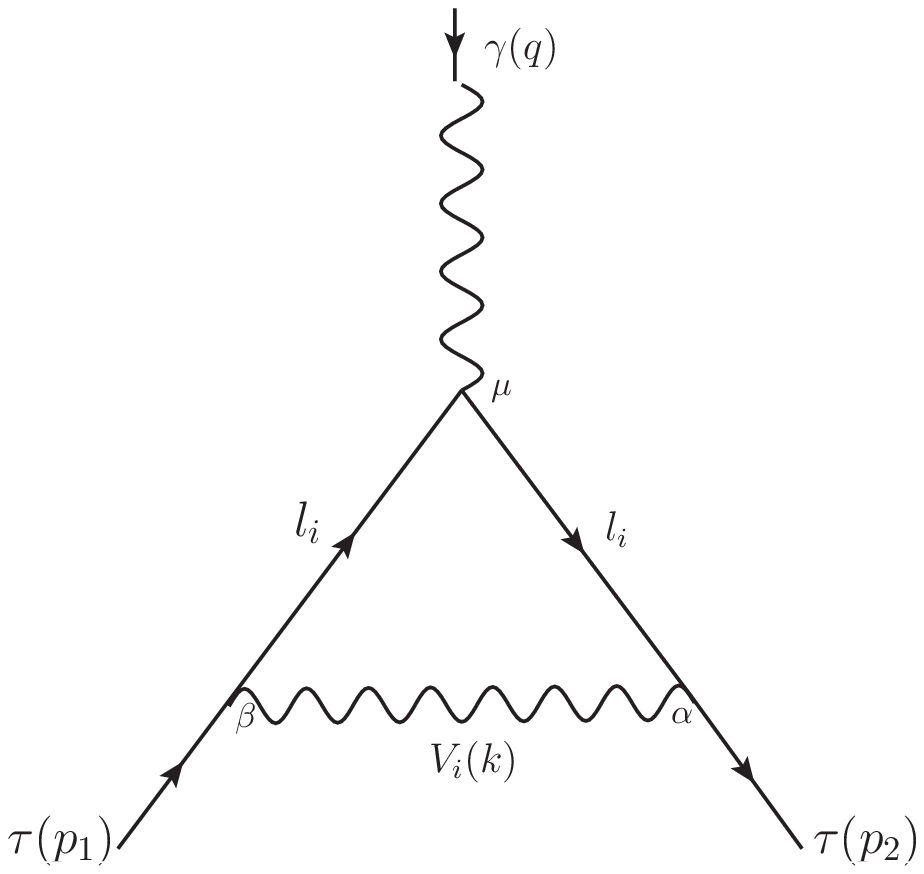}}  \\
\subfloat[]{\includegraphics[width=4.5cm]{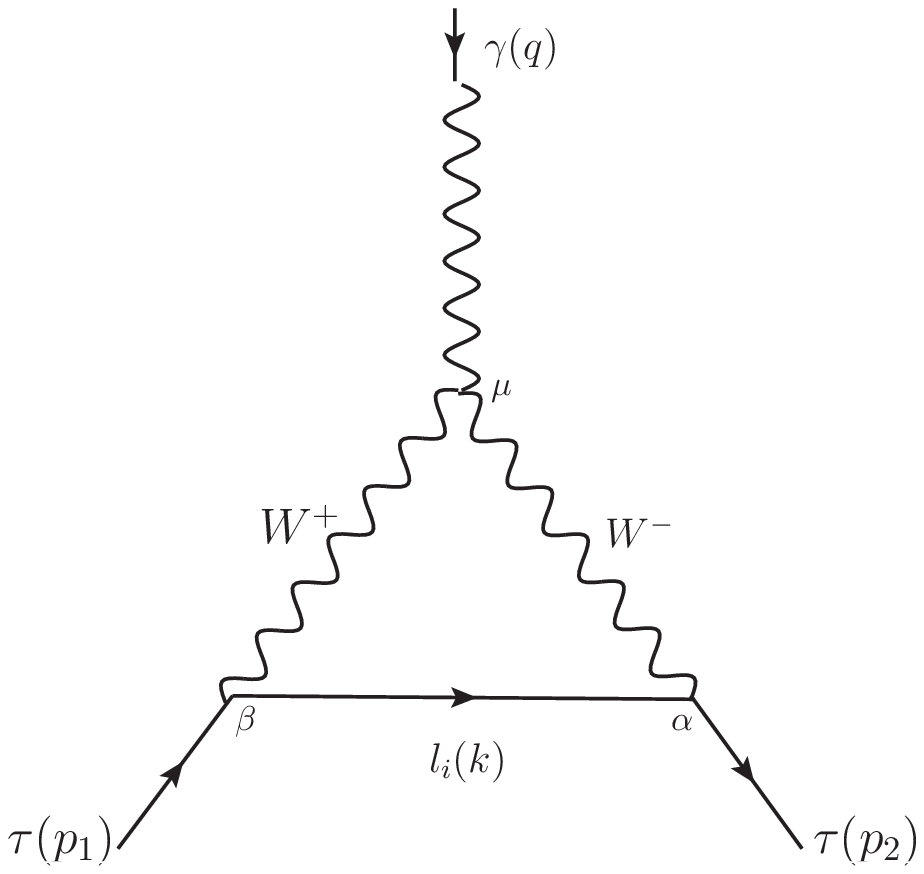}} 
\caption{ \label{dipoloMEgamma} Generic Feynman diagrams that contributes to the AMDM of the tau-lepton, $l_{i}\equiv \tau, \nu_{\tau}$.
 a) Scalar contribution, $S_{i}\equiv h_{0}$. b) and c) Vector contributions, $V_i \equiv \gamma, Z$. }
\end{figure}

\begin{figure}[H]
\center
\subfloat[]{\includegraphics[width=4.5cm]{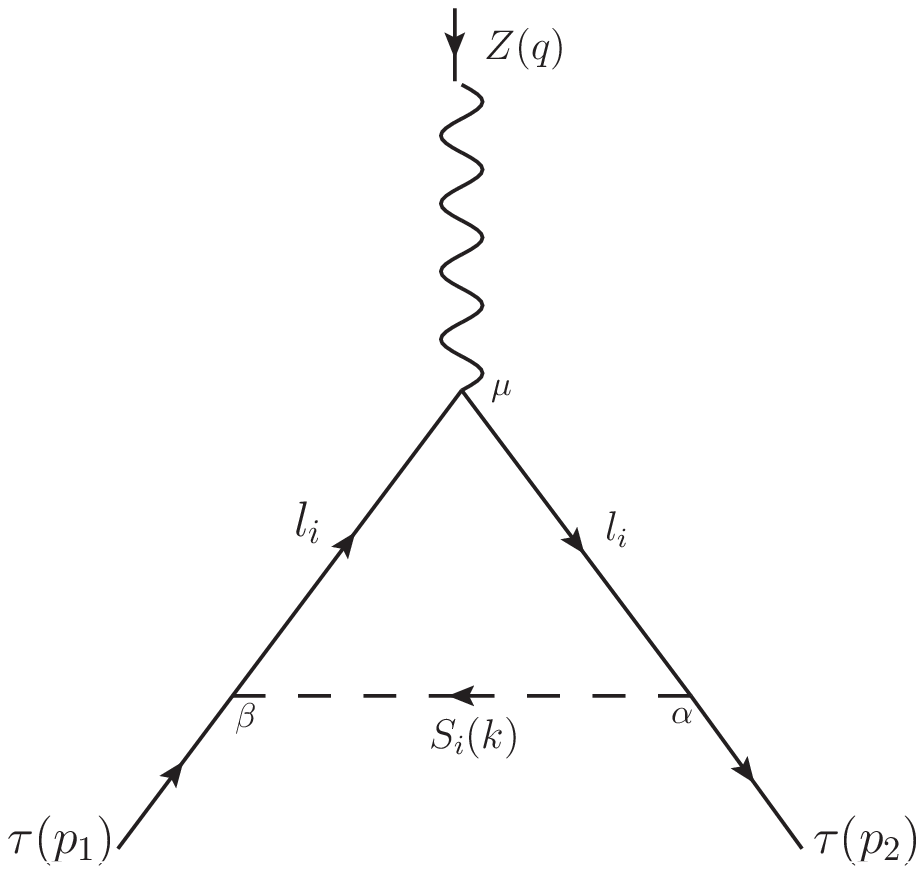}}
\subfloat[]{\includegraphics[width=4.5cm]{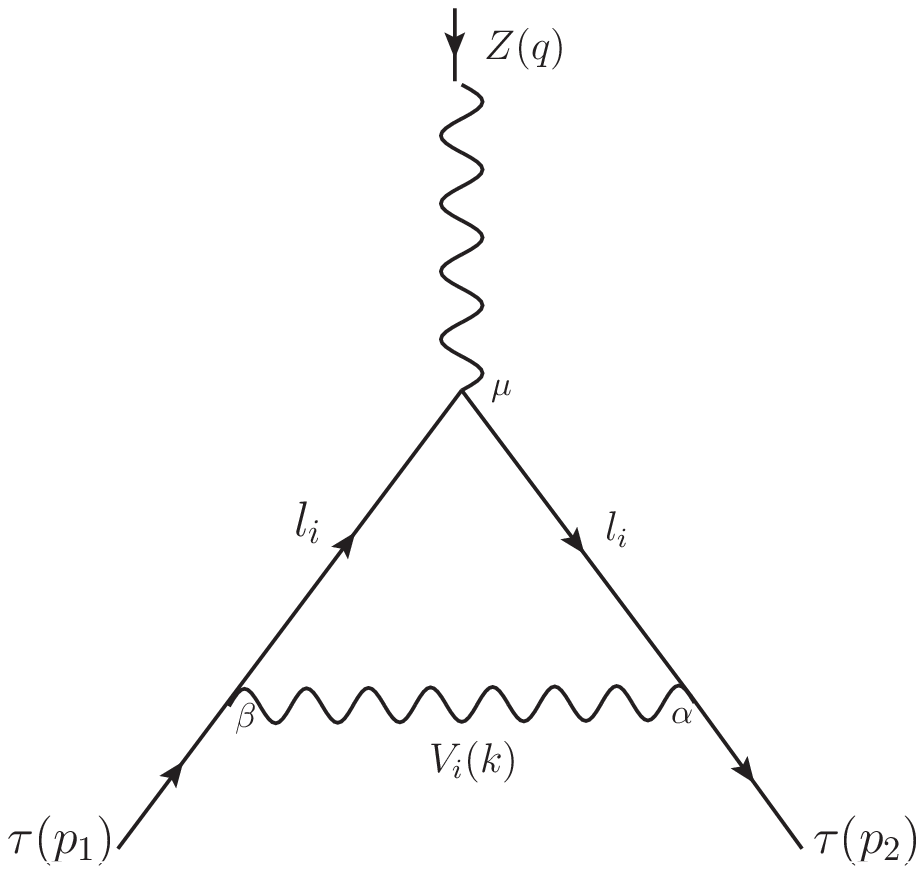}}  \\
\subfloat[]{\includegraphics[width=4.5cm]{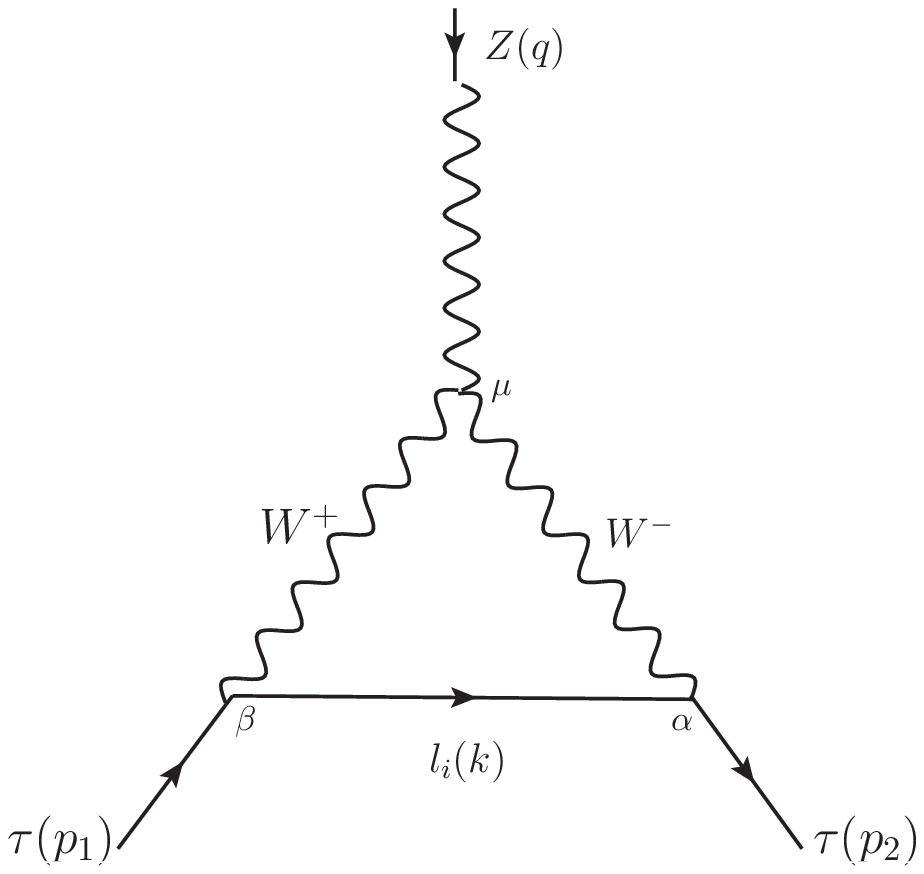}}
\subfloat[]{\includegraphics[width=4.5cm]{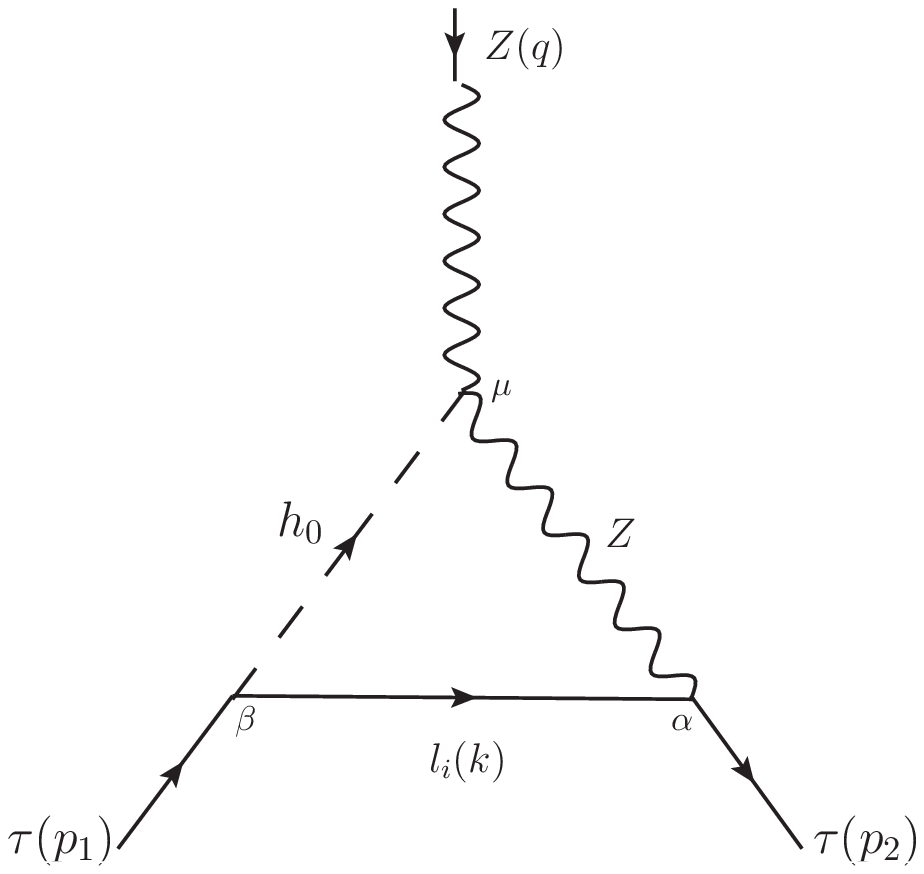}} 
\caption{ \label{dipoloMEZ} Generic Feynman diagrams that contributes to the AWMDM of the tau-lepton, $l_{i}\equiv \tau, \nu_{\tau}$.
 a) Scalar contribution, $S_{i}\equiv h_{0}$. b), c) and d) Vector contributions, $V_i \equiv \gamma, Z$.}
\end{figure}

 In the SM, we estimate at the one-loop level the contributions to the AMDM and AWMDM of the tau-lepton. These contributions are calculated in the unitarity gauge, so that the only Feynman diagrams that arise are those shown in Figs.~\ref{dipoloMEgamma} and~\ref{dipoloMEZ}. The first-order contributions for $ a^{SM}_{\tau} $ and $ a^{W-SM}_{\tau} $  are obtained from these figures.

The prediction of the AMDM of the $ \tau $-lepton in SM is calculated by considering only one-loop level contributions. In the literature these contributions found are usually catalogued as the quantum electrodynamics (QED) contribution without hadrons, and the electroweak contribution. 
In Table~\ref{atau-ME}, we provide the numerical values of the QED contribution and the partial electroweak contributions. In this table we can appreciate that indeed the QED contribution is the most important, followed by the electroweak contribution. 
Our result obtained on $ a^{SM}_{\tau} = 116090_\cdot 33\times 10^{-8}$, is comparable to that of Ref.~\cite{Eidelman:2007sb}.
Regarding the EDM of the tau-lepton, it is absent at this level.

\begin{table}[H]
\caption{Partial contributions to $ a^{SM}_\tau $. {\bf abc} denotes the different particles running in the loop of the vertex $\gamma \tau^{+} \tau^{-}$.
\label{atau-ME}}
\centering
\begin{tabular}{|c|c|}
\hline
 $\text{Couplings}\  \textbf{abc}$  & $\left(a^{SM}_\tau \right)^{ \textbf{abc} } $  \\
\hline
$ \gamma \tau \tau  $  & $ 116140_\cdot  97 \times 10^{-8}+ 0 \,i $  \\
\hline
$ Z \tau \tau $  & $ 51_\cdot 55 \times 10^{-8} + 0 \ i$  \\
\hline
$h_0\tau \tau $  & $ 0.09 \times 10^{-8} +0 \ i$  \\
\hline
$  \nu_\tau WW $  & $ -102_\cdot 29\times 10^{-8}+ 0 \ i$  \\
\hline
$ \textbf{Total} $  & ${\bf a^{SM}_{\tau} }= 116090_\cdot 33\times 10^{-8} +0\ i$  \\
\hline
\end{tabular}
\end{table}

We also estimate the AWMDM of the tau-lepton  with the initial and final particles in on-shell. The relevant diagrams in the unitary gauge are those shown in Fig.~\ref{dipoloMEZ}, and their numerical contributions are given in Table~\ref{awtau-ME}. In this table we can appreciate that the largest partial contribution to $ a^{W-SM}_\tau $, in absolute value, arises when $W^{+}$, $W^{-}$ and $\nu_\tau$ particles circulate in the loop.
The total contribution to $ a^{W-SM}_\tau $ is $-1_\cdot 9193\times 10^{-6} - 0_\cdot 5713\times 10^{-6}\ i$.
 Our result is comparable to that reported in Ref.~\cite{Bernabeu:1994wh}, although a slight difference prevails. This is due to the fact that we used current values  for the input parameters $m_Z$, $m_W$, $m_\tau$, $\sin \theta_W$ and $ \alpha $ (fine-structure constant).

\begin{table}[H]
\caption{Partial contributions to $ a^{W-SM}_\tau $. {\bf abc} denotes the different particles running in the loop of the vertex $Z \tau^{+} \tau^{-}$.
\label{awtau-ME}}
\centering
\begin{tabular}{|c|c|}
\hline
 $\text{Couplings}\  \textbf{abc}$  & $\left(a^{W-SM}_\tau \right)^{ \textbf{abc} } $  \\
\hline
$ \gamma \tau \tau  $  & $ 3_\cdot 09 \times 10^{-7}-1_\cdot 24 \times 10^{-7} \,i $  \\
\hline
$ Z \tau \tau $  & $ 4_\cdot 03\times 10^{-8} + 1_\cdot 86\times 10^{-8} \ i$  \\
\hline
$h_0\tau \tau $  & $ -6_\cdot 72\times 10^{-12} - 1_\cdot 38\times 10^{-11} \ i$  \\
\hline
$ W\nu_\tau \nu_\tau $  & $ -9_\cdot 14\times 10^{-7} -4_\cdot 66\times 10^{-7} \ i$  \\
\hline
$  \nu_\tau WW$  & $ -1_\cdot 37\times 10^{-6}+ 0 \ i$  \\
\hline
$ \tau h_0 Z  $  & $ 1_\cdot 35 \times 10^{-8} +0 \ i$  \\
\hline
$ \textbf{Total} $  & ${\bf a^{W-SM}_{\tau} }= -1_\cdot 92\times 10^{-6} - 0_\cdot 57\times 10^{-6}\ i$  \\
\hline
\end{tabular}
\end{table}

\newpage

\end{document}